\documentclass[12pt]{article}

\usepackage{times}
\usepackage{amsmath}
\usepackage{amsfonts}
\usepackage{amssymb}
\usepackage{graphicx}
\usepackage{soul}
\usepackage{hyperref}
\usepackage[english]{babel}
\usepackage[utf8]{inputenc}
\usepackage{caption}
\usepackage{xcolor}
\usepackage[noabbrev]{cleveref}
\usepackage[labelformat=parens,subrefformat=parens]{subfig}
\usepackage{siunitx}
\usepackage{xcolor}
\usepackage{layouts}
\usepackage{mathtools} \usepackage{tikz}
\usepackage{mwe}
\usepackage{bm}
\usepackage{breqn}
\usepackage{booktabs}
\usepackage{enumitem}
\usepackage[ruled,vlined]{algorithm2e} \usepackage[title]{appendix}

\newenvironment{myabstract}{\begin{quote} \bf}
{\end{quote}}

\begin{document}

\captionsetup[figure]{labelfont={bf},name={Fig.},labelsep=period}

\title{Cluster-based network modeling--automated robust modeling of complex
dynamical systems}

\author{
  Daniel Fernex,${}^{1}$, Bernd R. Noack${}^{2}$, Richard Semaan${}^{1\ast}$\\
  \normalsize{${}^{1}$Institut für Strömungsmechanik, Technische Universität Braunschweig,}\\
  \normalsize{Hermann-Blenk-Str. 37, 38108 Braunschweig, Germany}\\
  \normalsize{${}^{2}$Center for Turbulence Control, Harbin Institute of Technology,}\\
  \normalsize{Shenzhen 518058, People's Republic of China}\\
  \normalsize{$^\ast$To whom correspondence should be addressed; E-mail: r.semaan@tu-braunschweig.de.}
}

\date{}
 
\maketitle

\begin{myabstract}
        We propose a universal method 
        for data-driven modeling of complex nonlinear dynamics
        from time-resolved snapshot data without prior knowledge.
	Complex nonlinear dynamics govern many fields of science and engineering.
	Data-driven dynamic modeling often assumes 
	a low-dimensional subspace or manifold for the state.
	We liberate ourselves from this assumption by 
	proposing cluster-based network modeling (CNM) 
	bridging machine learning, network science, and statistical physics.
	CNM only assumes smoothness of the dynamics in the state space,
	robustly describes short- and long-term behavior
	and is fully automatable as it does not rely on application-specific knowledge.
	CNM is demonstrated for the Lorenz attractor, 
	ECG heartbeat signals, 
	Kolmogorov flow, and a high-dimensional actuated turbulent boundary layer.
	Even the notoriously difficult modeling benchmark 
	of rare events in the Kolmogorov flow is solved.
	This automatable universal  data-driven representation 
	of complex nonlinear dynamics  
	complements and expands network connectivity science and
	promises new fast-track avenues
	to understand, estimate, predict and control complex systems
        in all scientific fields.
\end{myabstract}
 
\section{Introduction}

Climate, epidemiology, brain activity, financial markets
and turbulence constitute examples of complex systems.
They are characterized by a large range of time and spatial scales,
intrinsic high dimensionality and nonlinear dynamics.
Dynamic modeling for the long-term features is a key enabler for understanding,
state estimation from limited sensors signals, 
prediction, control, and optimization.
Data-driven modeling has made tremendous progress in the last decades,
driven by algorithmic advances, accessibility to large data, and hardware speedups.
Typically, the modeling is based on a low-dimensional
approximation of the state and system identification in that approximation.

The low dimensional approximation may be achieved with subspace modeling methods, 
such as proper orthogonal decomposition  (POD)
models~\cite{Holmes2012Book,Benner2015SIAM}, dynamic mode decomposition (DMD)~\cite{Tu2014JCD} 
and empirical dynamical modeling~\cite{Ye2015PNAS}, to name only a few.
Autoencoders \cite{Brunton2020arfm} represent a general nonlinear dimension reduction to a low-dimensional feature space.
The dynamic system identification is significantly simplified in this feature space. 

An early breakthrough in system identification was reported by Bongard and
Lipson~\cite{Bongard2007PNAS} using symbolic regression.
The method performs a heuristic search of the best equation that describes the dynamics~\cite{Schmidt2009science}. 
They are however expensive and not easily scalable to large systems. 
Recent developments in parsimonious modeling lead to
the ``sparse identification of nonlinear dynamics'' (SINDy) algorithm that identifies 
accurate parsimonious models from data~\cite{Brunton2016PNAS}.
Similarly, SINDy is not easily scalable to large problems. 
The computational expense becomes exorbitant already 
for moderate dimensional feature spaces.

This limitation may be by-passed by black-box techniques.
These include Volterra series~\cite{Fu1973IJC}, autoregressive
models~\cite{Chatfield2000Book}
(e.g., ARX, ARMA, and NARMAX), eigensystem realization algorithm
(ERA)~\cite{Juang1994Book},
and neural network (NN) models~\cite{Wang2015IEEE}.
These approaches, however, have limited interpretability and provide little physical insights.
Some (e.g. NN) require large volumes of data and long training time, 
luxuries that are not always at hand. 

In this study, we follow a novel modeling paradigm
starting with a time-resolved snapshot set.
We only assume smoothness of the dynamics in the state space
liberating ourselves from the requirement of a low-dimensional subspace or manifold for the data
and analytical simplicity of the dynamical system.
The snapshots are coarse-grained
into a small number of centroids with clustering.
The dynamics is described by a network model with continuous transitions between the centroids.
The resulting cluster-based network modeling (CNM) 
uses time-delay embedding
to identify models with an arbitrary degree of complexity and nonlinearity.
The methodology is developed 
within the network science~\cite{Newman2008PT,Barabasi1999Science,Barabasi2003SA}
and statistical physics~\cite{Norris1998} frameworks.
Due to its generic nature, network analysis is being increasingly used to investigate complex
systems \cite{Marwan2009PLA,Taira2016JFM}.
The proposed method builds on previous work by Kaiser et al.~\cite{Kaiser2014JFM},
where clustering is used to coarse-grain the data into representative states and the
temporal evolution is modeled as a probabilistic Markov model.
By construction, 
the state vector of cluster probabilities converges 
to a fixed point representing the post-transient attractor,
i.e., the dynamics disappear.
A recent improvement~\cite{Li2020JFM} models
the transition dynamics between the network nodes 
as straight constant-velocity `flights'
with a travel time directly inferred from the data.
The present study expands on these innovations 
and generalizes the approach to arbitrary high-order chains
with time-delay coordinates~\cite{Ching2013Book},
and introduces a control-oriented extension to include external inputs and control.
Besides its accuracy, one major advantage the method has is the ability to control the resolution level through adaptive coarse-graining.

Dynamics of complex systems is often driven by complicated small-scale (sometimes microscopic) interactions 
(e.g. turbulence, biological signaling) that are either unknown or very expensive
to fully-resolve~\cite{Daniels2015NC}.
The resolution of cluster-based network modeling can be adapted to match any desired level, 
even when microscopic details are not known. 
This universal representation of strongly nonlinear dynamics, enabled by adaptive coarse-graining and a probabilistic foundation, 
promises to revolutionize our ability to understand, estimate, predict and control complex systems in all scientific fields. 
The method is inherently robust and honest to the data.
It requires no assumption on the analytical structure of the model,
and is computationally tractable, even for high-degrees of freedom problems. 
A code is available at: \url{https://github.com/fernexda/cnm}.
 
\section{Cluster-based network modeling}

\begin{figure*}[h!]
  \centering
  \includegraphics[width=\textwidth]{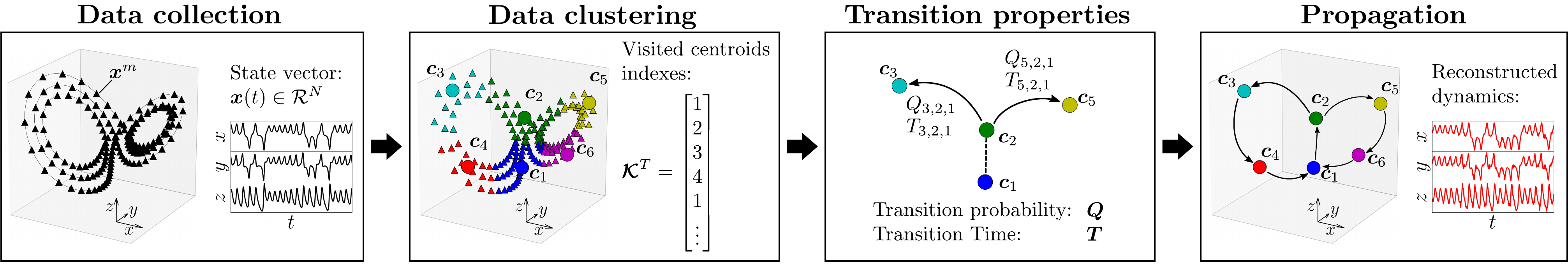}
  \caption{
    \textbf{Cluster-based network modeling methodology.} $M$ consecutive $N-$dimensional states $\bm{x}(t)\in\mathcal{R}^{N\times M}$ are
      collected at fixed sampling frequency. 
Based on their similarity, the states are grouped into $K$ clusters. The
      network nodes are computed as the cluster centroids $\bm{c}_i$, and the transition
      time $\bm{T}$ and transition probability $\bm{Q}$ between the nodes are
      identified from the data.
The CNM dynamics are propagated as consecutive flights between centroids.
      Each transition is characterized by its destination, given by $\bm{Q}$,
      and its transit time given by $\bm{T}$.
    \label{Fig:Meth:CNM}
  }
\end{figure*}

Robust probability-based data-driven dynamical modeling for complex nonlinear systems 
has the potential to revolutionize our ability to predict and control these systems.
Cluster-based network models (CNM) reproduce the dynamics on a directed network, 
where the nodes are the coarse-grained states
of the system. 
The transition properties between the nodes are based on
high-order direct transition probabilities identified from the data.
The model methodology is applied to a variety of dynamical systems, from canonical
problems such as the Lorenz attractor to rare events to high degrees of freedom systems such as a
boundary layer flow simulation.
The general methodology is illustrated in Fig. \ref{Fig:Meth:CNM} with the
Lorenz system and is detailed in the following.
\paragraph{Data collection and clustering.}
The starting point of CNM
is the data collection of $M$ consecutive discrete $N-$dimensional state of the system
$\bm{x}(t)\in \mathcal{R}^{N}$ equally-spaced in time with $\Delta t$, such that the state at $t^m$ is
$\bm{x}(t^m)=\bm{x}(m\Delta t)=[x_1^m,\ldots,x_N^m]$.
The discrete states are grouped into $K$ clusters $\mathcal{C}_k$ and the network nodes are identified as the clusters' centroids $\bm{c}_k$, defined as the average of the states
in each cluster. In this study, clustering is achieved with the unsupervised 
$k-$means++ algorithm~\cite{David2006,Jain1999ACMCS} that minimizes the inner-cluster variance.
Other clustering algorithms are possible.
The choice is a problem-dependent option.
The vector $\bm{\mathcal{K}} = [\mathcal{K}_1, \ldots, \mathcal{K}_ I]$,
$\mathcal{K}_i \in [1,K]$, contains the indexes of the consecutively-visited clusters over the entire time
sequence, such that $\mathcal{K}_i$ is the index of the $i$th visited cluster. The first and last clusters
are $\mathcal{C}_{\mathcal{K}_1}$ and $\mathcal{C}_{\mathcal{K}_I}$, respectively.
The size $I$ of $\bm{\mathcal{K}}$ is equal to the number of transitions between
$K$ centroids over the entire ensemble plus one.
We note that two sequential cluster visits are not necessarily equally-spaced in time,
but rather depend on the state's rate of change in their vicinity.

\paragraph{Transition properties.}
Before we detail the transition properties of cluster-based network models~\cite{Li2020JFM} ,
we briefly review those of cluster-based Markov models~\cite{Kaiser2014JFM} 
upon which the current method builds.
In cluster-based Markov models, the state variable is the cluster population 
$\bm{p}=\left[ p_1,\ldots,p_K \right ]^{\mathrm{T}}$, 
where $p_i$ represents the probability to be in cluster $i$
and the superscript $\mathrm{T}$ denotes the transpose.
The transitions between clusters are modeled with a first-order Markov model. 
The probability to move from cluster $\mathcal{C}_j$ to cluster $\mathcal{C}_k$ is described
by the transition matrix $\bm{P}=(P_{k,j}) \in {\cal R}^{K \times K}$ as
\begin{equation}
  P_{k,j}=\mathrm{Pr}\left(\mathcal{K}_i=k\lvert \mathcal{K}_{i-1}=j\right)\,.
  \label{Eqn:TransMat}
\end{equation}
The transition matrix $\bm{P}$ is computed as
\begin{equation}
	P_{k,j}=\frac{n_{k,j}}{n_{j}}\,,
\end{equation}
where $n_{k,j}$ are the number of samples that move from $\mathcal{C}_{j}$
to $\mathcal{C}_k$, and $n_{j}$ is the number of transitions departing from $\mathcal{C}_{j}$ regardless of the destination point.

In~\cite{Kaiser2014JFM}, the transition time $\Delta t$ is a user-specified constant.
Let $\bm{p}^l$ be the probability vector at time $t^l = l \Delta t$,
then the change in one time step is described by 
\begin{equation}
  \bm{p}^{l+1} = {{{\bm{P}}}} \> \bm{p}^l\,.
  \label{Eqn:CMM}
\end{equation}
With time evolution, equation~\eqref{Eqn:CMM}
converges to the asymptotic probability 
$\bm{p}^{\infty} := \lim\limits_{l\to \infty} \bm{p}^l$.
In a typical case, equation \eqref{Eqn:CMM} has a single fixed point $\bm{p}^{\infty}$.

Conversely, CNM relies on the \emph{direct transition matrix} $\bm{Q}$,
which ignores inner-cluster residence probability and only considers 
inter-cluster transitions.
The direct transition probability is inferred from data as
\begin{equation}
  Q_{k,j}=\frac{n_{k,j}}{n_{j}}\,,
\end{equation}
with $Q_{j,j}=\mathrm{Pr}(\mathcal{K}_i=j|\mathcal{K}_{i-1}=j)=0$,
by the very definition of a direct transition.
Generalizing to an $L-$order model,
which is equivalent to using time-delay coordinates, the direct transition probability is expressed as
$\mathrm{Pr}\left(\mathcal{K}_i\lvert\mathcal{K}_{i-1},\ldots,\mathcal{K}_{i-L}\right)$.
Illustrating for a second-order model the probability to move to $\mathcal{C}_l$ having previously
visited $\mathcal{C}_k$ and $\mathcal{C}_j$ is given by
\begin{equation}
  \label{Eqn:HOMM}
  Q_{l,k,j}=\mathrm{Pr}(\mathcal{K}_{i}=l\lvert\mathcal{K}_{i-1}=k,\mathcal{K}_{i-2}=j)\,.
\end{equation}
Time-delay embedding is a cornerstone of dynamical systems~\cite{Takens1981}.
The optimal Markov chain order
$L$ is problem-dependent (see Appendix \ref{Sec:ParamSelection}). Larger $L$ values are typically necessary for problems with
complex phase-space trajectories.
In this study, we shall demonstrate how time-delay embedding benefits extend to higher-order cluster-based network models.

\begin{figure}[h!]
  \centering
  \includegraphics[width=0.4\textwidth]{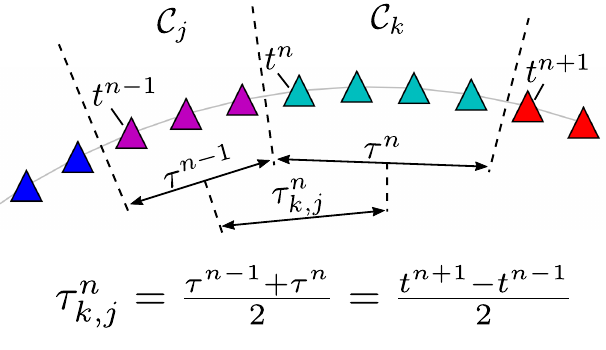}
  \caption{
    \textbf{Definition of the transition time between clusters
      $\mathcal{C}_j$ and $\mathcal{C}_k$.}
      The transit time $\tau^{n}$ in $\mathcal{C}_k$ at iteration $n$ is the time range spanned by the data
      entry and exit times in the clusters, $t^{n}$ and $t^{n+1}$.
      The individual transition time $\tau_{k,j}^n$ is defined as the average transit time between two neighboring
      clusters.
    \label{Fig:Meth:Time}
  }
\end{figure}

The second transition property is the transition time.
For Markov models, the time step is a critical user-defined design parameter.
If the time step is too small, the cluster-based Markov model idles many times in each cluster for a stochastic number of times before transitioning to the next cluster. 
The model-based transition time may thus significantly deviate from the deterministic data-driven trajectories through the clusters. 
If the time step is too large, one may miss intermediate clusters. 
This design parameter can be avoided in cluster-based network modeling (CNM). 
The key idea is to abandon the `stroboscopic’ view and focus on non-trivial transitions,
thus avoiding rapid state diffusion.
Let $t^n$ and $t^{n+1}$ be the time of the first and last snapshots to enter and, respectively, to leave
$\mathcal{C}_k$ at the $n$th iteration (Fig. \ref{Fig:Meth:Time}).
Here, iterations refer to the sequential jumps between the centroids.
The residence time $\tau^n=t^{n+1}-t^{n}$  corresponds to the duration of the state
transit in cluster $\mathcal{C}_k$ at this iteration.
We define the individual transition time from cluster $j$ to cluster $k$ for one iteration as half the residence time 
of both clusters, 
\begin{equation}
  \tau_{k,j}^n=\frac{\tau^{n-1}+\tau^n}{2}=\frac{t^{n+1}-t^{n-1}}{2}\,.
\end{equation}
Averaging all $n_{k,j}$ individual transition times
between $\mathcal{C}_{j}$ to $\mathcal{C}_{k}$ 
yields the transition time $T_{k,j}=1/n_{k,j}\sum_{n=1}^{n_{k,j}}\tau_{k,j}^n$.
This definition may appear arbitrary but is the least-biased guess consistent with the available data.
Similar to the direct transition matrix $\bm{Q}$ for an $L-$order chain, 
the transition time matrix $\bm{T}=\left(T_{k,j}\right)\in\mathcal{R}^{K\times K}$ 
also depends on the $L-1$ previously visited centroids.
When $L$ is large, this could yield to two storage-intensive
$L+1-$dimensional tensors $\bm{Q}$ and $\bm{T}$ with $K^{L+1}$ elements.
The expensive tensor creation and storage is circumvented by a lookup
table (LUT), where only non-zero entries that correspond to actual transitions
are retained. 
The look-up tables are typically orders-of-magnitude smaller than the full tensors.
(see Appendix \ref{Sec:CNMMethod}).

\paragraph{Propagation.}
The final step in cluster-based network modeling propagates the state motion.
We assume a uniform state propagation between two centroids $\bm{c}_{j}$ and $\bm{c}_{k}$ as,
\begin{equation}
  \bm{x}(t) = \alpha_{kj} (t) \bm{c}_{k} + [1 - \alpha_{kj}(t)] \bm{c}_{j},\, \quad
  \alpha_{kj}=\frac{t_{j}-t}{T_{k,j}}\,,
\end{equation}
where $t_{j}$ is the time when the centroids $\bm{c}_{j}$ is left.
The motion between the centroids may be interpolated with splines for
smoother trajectories.
As CNM is purely data-driven, the model quality is directly
related to that of the training data. 
More specifically, the sampling frequency and total
time range must be selected, such that all relevant dynamics are 
captured and are statistically fully converged.
This usually requires a larger amount of data than other data-driven methods,
such as ARMA and SINDy.

\section{Results}

\paragraph{CNM of the Lorenz system.}

\begin{figure*}
  \centering
  \includegraphics[width=\textwidth]{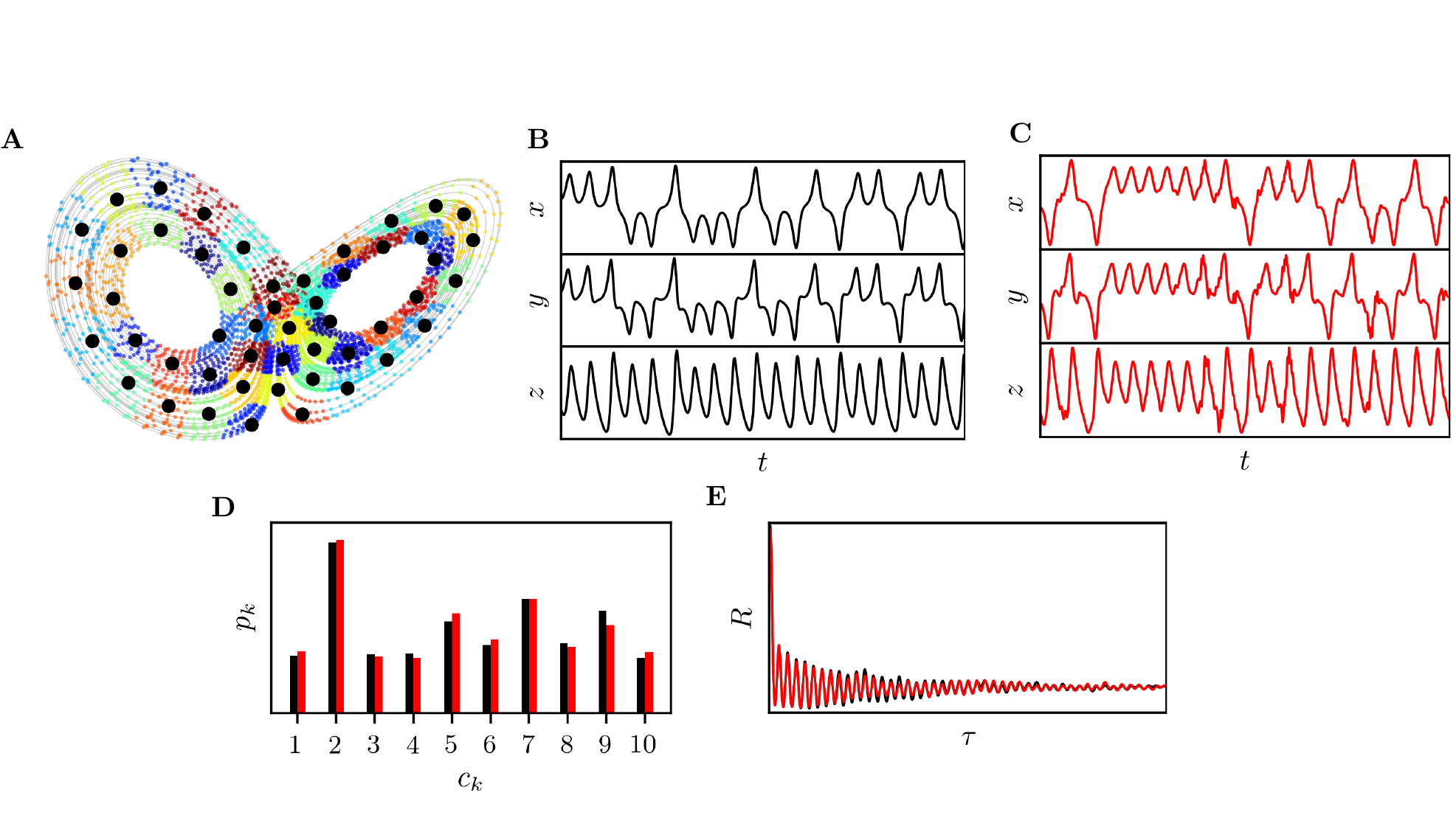}
  \caption{
    \textbf{Cluster-based network modeling of the Lorenz system.}
    (\textbf{A}) Phase-space representation of the data clustering. The
    centroids are depicted with black circles and the small circles are the
    snapshots, colored by their cluster affiliation.
    The CNM accuracy is demonstrated in the accurate reproduction of 
    (\textbf{B})-(\textbf{C}) the time series, (\textbf{D})  the cluster probability distribution,
    and (\textbf{E}) the autocorrelation function.
    Black and red coloring denotes the reference and CNM data, respectively.
\label{Fig:LorenzSOC:Summary}
}
\end{figure*}

CNM is applied to the Lorenz system, a widely-used canonical chaotic dynamical system~\cite{Lorenz1963JAS}
defined by three coupled nonlinear differential equations,
\begin{align}
  \frac{\mathrm{d} x}{\mathrm{d} t} &=\sigma (y-x) \nonumber \\
  \frac{\mathrm{d} y}{\mathrm{d} t} &=x (\rho-z) -y \\
  \frac{\mathrm{d} z}{\mathrm{d} t} &=x y-\beta z\,, \nonumber
  \label{Eqn:LorenzSystem}
\end{align}
where the system parameters are here defined as $\sigma=10$, $\rho=28$ and
$\beta=8/3$.
The data are clustered with $K=50$ centroids, depicted in Fig.
\ref{Fig:LorenzSOC:Summary}A. The snapshots are colored based on their cluster
affiliations.
CNM is performed with a chain order $L=22$ using $\approx 17000$ transitions, which
cover the same time range as that of the original data.
The optimal $K$ and $L$ values are problem-dependent.
They are identified for the Lorenz system through a parametric study,
where the root-mean square error of the autocorrelation function between the
reference data and the model is minimized (c.f. Appendix
\ref{Sec:ParamSelection}).

Time series obtained with CNM agree very well with the reference data (Fig.
\ref{Fig:LorenzSOC:Summary}B and
C). 
The oscillating amplitude growth in both ears, as well as the ear
switching, are correctly captured.
The cluster probability distribution (CPD) $q_k, \, k=1,\ldots,K$ provides the probability of the state
to be in a specific cluster. It indicates whether the modeled trajectories populate the phase space similarly
to the reference data (c.f. Appendix \ref{Sec:Validation}). The CPD for both the data and CNM is shown in Fig.
\ref{Fig:LorenzSOC:Summary}D. For clarity, $q_k$ is shown with 10 clusters only
instead of the full 50 clusters.
As the figure shows, CNM accurately reproduces the probability distribution. 
Following Protas et al.~\cite{Protas2015JFM}, the cluster-based network model is validated based on the autocorrelation function of the state vector.
This function avoids the problem of comparing two trajectories with finite dynamic prediction horizons due to phase mismatch. 
 The autocorrelation function also yields the
fluctuation energy at vanishing delay $R(0)$ and can be used to infer the
spectral behavior (see Appendix \ref{Sec:Validation}).
As Fig. \ref{Fig:LorenzSOC:Summary}E shows, CNM accurately reproduces the fast oscillatory decay, even after dozens of oscillations,
as well as the fluctuation energy $R(0)$, which is reproduced with a 2.8\% rms error.
This performance is in contrast to the cluster-based Markov models,
where time integration leads to the average flow, 
and to first-order cluster-based network models~\cite{Li2020JFM},
where the prediction accuracy is significantly lower.
A detailed comparison between the cluster-based Markov model, the first-order cluster-based network model, 
and the current model is provided in Appendix \ref{Sec:CNMUpgrade}.

\paragraph{Demonstration on examples.}

Cluster-based network modeling is applied to numerous examples, ranging from analytical systems 
to real-life problems using experimental and simulation data. The main
results are summarized in Fig.  \ref{Fig:Res:SummaryFig}. 
Details on each application are provided in Appendix \ref{Sec:SM:Results}.
\begin{figure*}[h!]
  \centering
  \includegraphics[width=\textwidth]{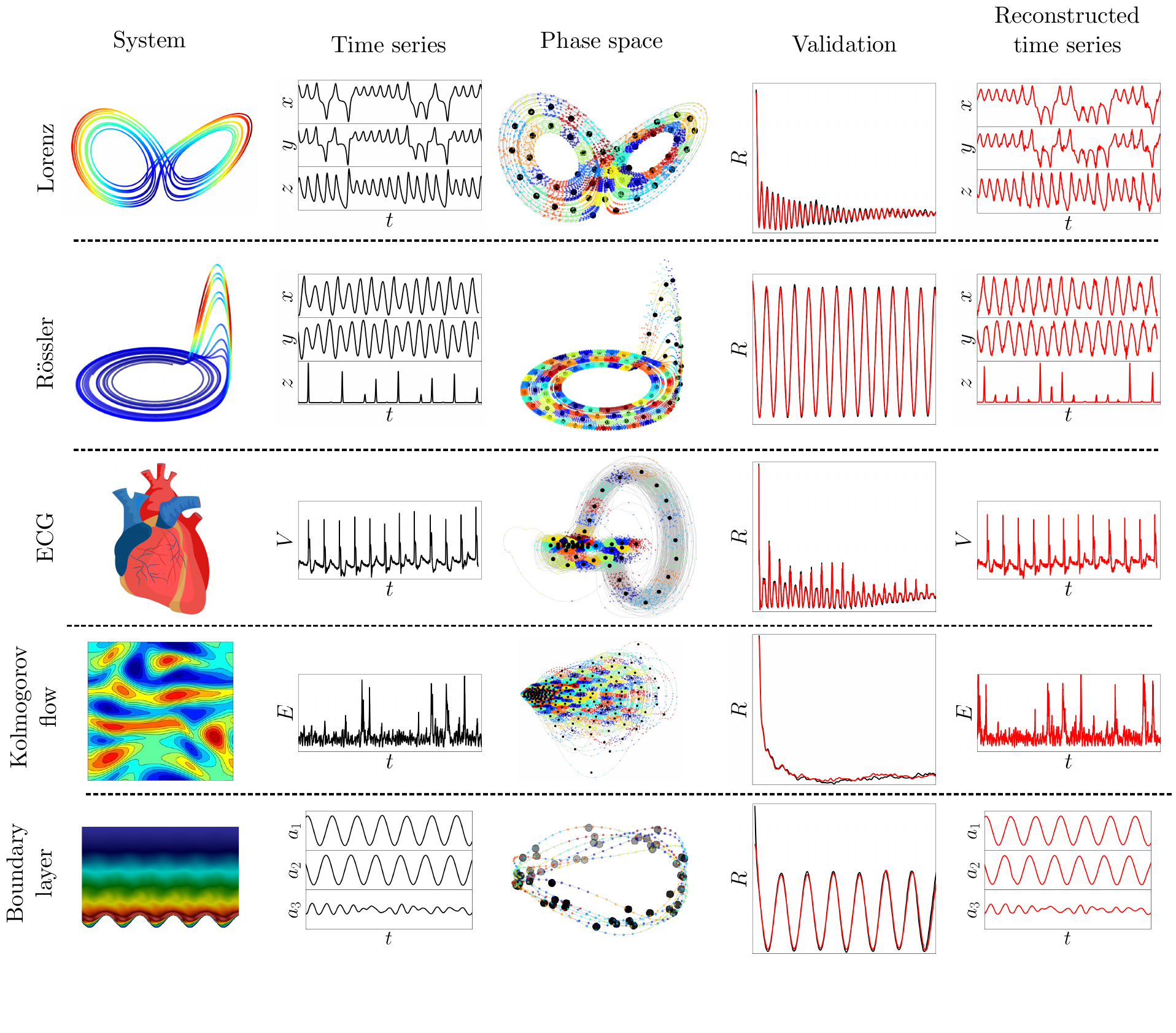}
  \caption{
    \textbf{The cluster-based network modeling implemented on five applications covering a
    wide range of dynamics.}
    The first two applications are three-dimensional chaotic systems, the Lorenz
    and Rössler attractors. The two following examples are one-dimensional
    experimental measurements from an electrocardiogram and numerical simulation
    of the dissipation energy in a Kolmogorov flow. The final application is a
    large-eddy simulation of an actuated turbulent boundary layer.
The excellent match of the autocorrelation functions for all applications
    demonstrate the CNM's ability to capture the relevant dynamics 
    for any complex nonlinear system.
The modeled time series faithfully reconstruct the data including
    the intermittent quasi-random bursts of the Kolmogorov dissipation energy,
    as well as the $z-$component pulses of the Rössler system.
    \label{Fig:Res:SummaryFig}
  }
\end{figure*}
The first two applications are the Lorenz~\cite{Lorenz1963JAS} and Rössler~\cite{Rossler1976PLA} attractors, 
typical candidates for dynamical systems analysis.
The two systems are governed by simple equations
and exhibit chaotic behavior under specific parameter values.
The following two implementations are one-dimensional systems: electrocardiogram
measurements (ECG)~\cite{Brunton2017NC}, and the dissipative energy from a
Kolmogorov flow~\cite{Farazmand2017SA}.
Whereas the ECG exhibits the regular heartbeat pattern, the
dissipative energy of the Kolmogorov flow is quasi-random with intermittent bursts. 
The last CNM application is a high-dimensional large eddy simulation of an
actuated turbulent boundary layer for skin friction reduction~\cite{Albers2020FTaC}.
The clustering step on this $\approx 5$ million grid cells simulation is performed on the mode coefficients of a lossless proper
orthogonal decomposition.
This dimensionality reduction step significantly reduces the computational load while yielding the same clustering outcome as the full difference matrix~\cite{Kaiser2014JFM}. 
The boundary layer time series are therefore represented with the mode coefficients.

In each example, both the qualitative and quantitative dynamics are faithfully captured.
The reconstructed time series are hardly distinguishable from the original data.
Intermittent events such as the peaks in the Rössler $z-$component and the
dissipation energy bursts of the Kolmogorov flow are statistically very well
reproduced.
The autocorrelation distributions of both reference data and models 
match perfectly over the entire range, demonstrating both robustness
and accuracy.
We note that robustness is inherent to CNM, since the modeled state
always remains close to the training data.

\begin{figure}[h!]
  \begin{center}
    \includegraphics[width=11cm]{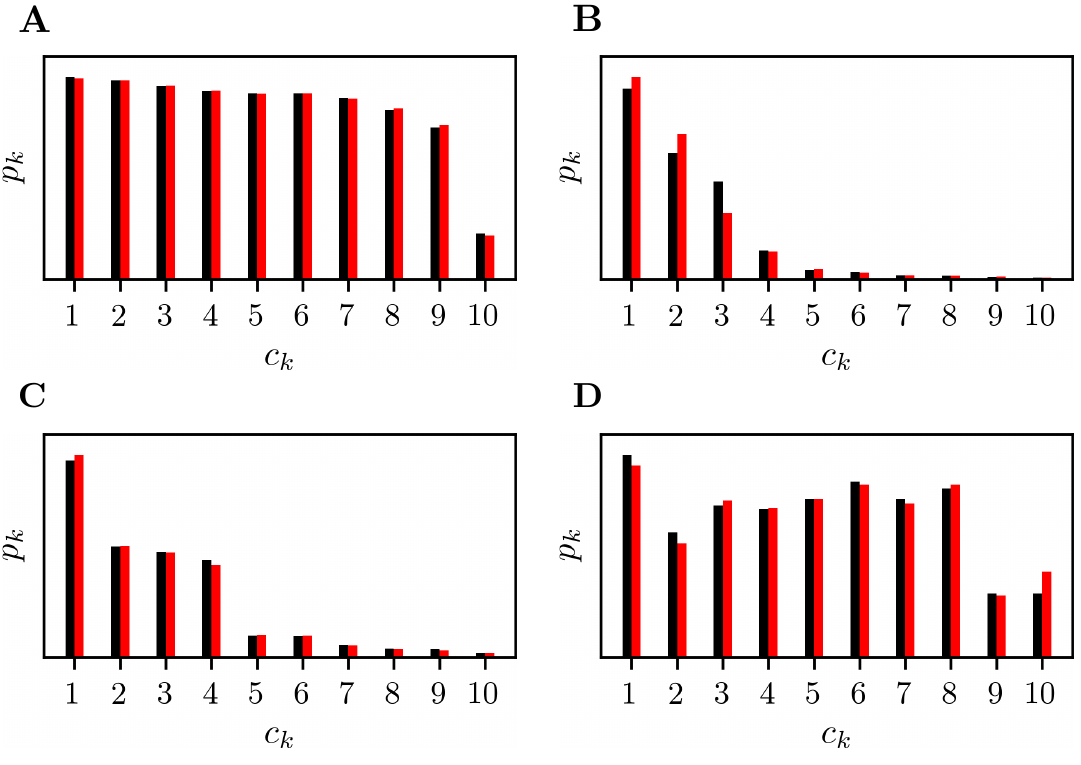}
  \end{center}
  \caption{
    \textbf{Cluster probability distribution (CPD) of the data and CNM for four applications.}
    (\textbf{A})-(\textbf{D}) CPD of the Rössler system, ECG
    signal, Kolmogorov flow dissipation energy, and actuated turbulent boundary layer, respectively.
    For all cases, the data (black) and CNM (red) are in good agreement. The
    specific features of each dataset, such as the rare events of the
    Kolmogorov dissipation energy and the fast heartbeat pulses are
    probabilistically well reconstructed by CNM.
  }
  \label{Fig:SummaryCPV}
\end{figure}

The CPD of the data and CNM for the Rössler system, the ECG signal, the Kolmogorov flow dissipation energy and
the actuated turbulent boundary layer are presented in Fig. \ref{Fig:SummaryCPV}.
For all cases, CNM accurately reconstructs the distributions. Remarkably,
the probabilities of less visited-clusters corresponding to rare events for the
Kolmogorov flow (Fig. \ref{Fig:SummaryCPV}C) or fast events such as the peaks
in the $z$ directions of the Rössler attractor (Fig. \ref{Fig:SummaryCPV}A) and
the heartbeat pulse (Fig. \ref{Fig:SummaryCPV}B) are very well captured by CNM.

\begin{figure}
  \centering
  \includegraphics[width=15.5cm]{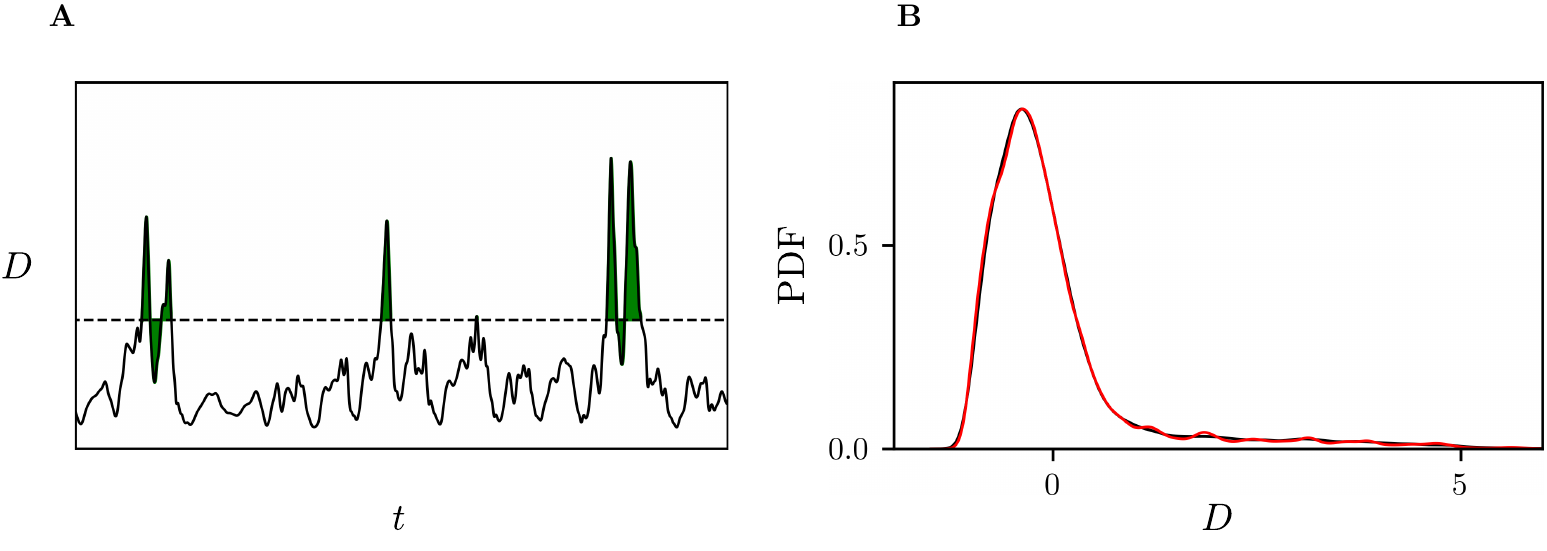}
    \caption{
  	\textbf{
  		Rare events from the Kolmogorov flow dissipation energy.
  	}
  	(\textbf{A}) Time series of the dissipation energy $D$. The dashed line
  	denotes an arbitrary threshold beyond which the peaks, represented with
  	green filling, are considered a burst.
  	(\textbf{B}) Probability distribution of the data (black) and CNM (red).
  	Both the main peak and the decaying tail of the distribution are accurately
  	reproduced.
  	\label{Fig:Res:RareEvents}
  }
\end{figure}

A special characteristic of CNM is its ability to accurately model and predict systems with rare events.
This ability is rooted in the probabilistic framework upon which CNM is constructed,
where the recurrence properties are the same as the reference data.
If one cluster is visited multiple times (or seldom) in the data, it will also be a recurrence point of the CNM.
A generic example of a rare events problem is the Kolmogorov flow~\cite{Wan2018PLOS}, a two-dimensional
incompressible flow with sinusoidal forcing. With a sufficiently high forcing
wavenumber, the flow becomes unstable and the dissipation energy $D$ exhibits
intermittent and spontaneous bursts (c.f. Fig. \ref{Fig:Res:RareEvents}A). The
dashed line denotes an arbitrary threshold beyond which a peak is considered a
rare event.
The probability distribution function (PDF) of the dissipation energy from the data and CNM are compared in Fig.
\ref{Fig:Res:RareEvents}B.
The main peak centered around zero
reflect the stochastic nature of the dissipation energy,
whereas the tail depicts rare events
whose occurrence probability decreases with their amplitude.
As the figure shows, CNM accurately captures the probabilistic behavior of the dissipation
energy. Both the main stochastic peak and the rare event tail of the
distribution are well reproduced.
Moreover, the total number of bursts in the current sequence is well reproduced,
with 58 bursts in the original data compared to 62 for CNM.

\paragraph{Control-oriented cluster-based network modeling (CNMc).}

\begin{figure}[h!]
  \begin{center}
    \includegraphics[width=14cm]{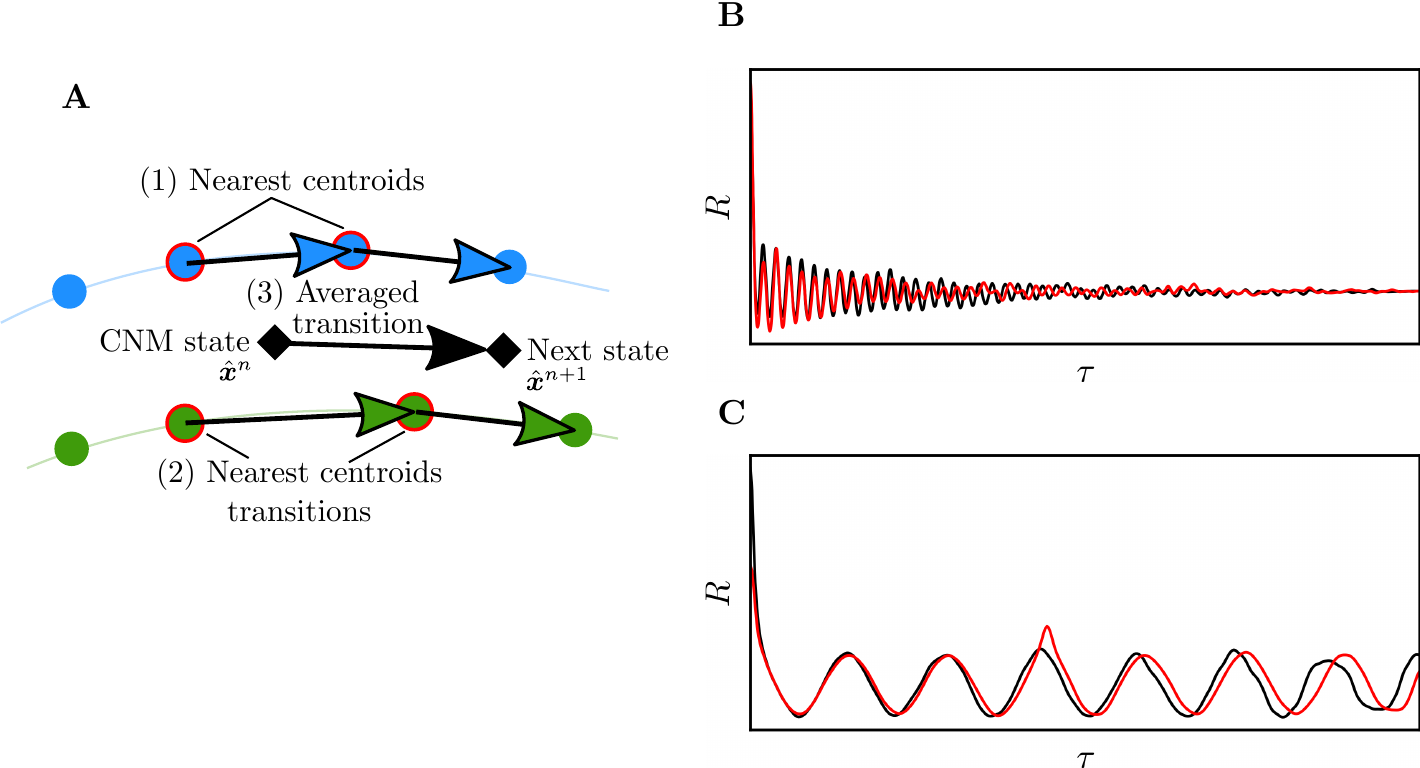}
  \end{center}
  \caption{
    \textbf{Control-oriented cluster-based network modeling (CNMc).}
     (\textbf{A}) CNMc iteratively propagates the state in the phase-space populated with the centroids from the two operating conditions with the closest control parameters.
(1) Neighboring centroids to the current state $\bm{\hat{x}}^n$ at iteration $n$ are first identified.
    (2) Their transition properties are calculated and then
    (3) averaged to determine the next 
    state $\bm{\hat{x}}^{n+1}$.
    CNMc accuracy is demonstrated by the autocorrelation function distributions of the data (black) and the predicted case (red) for the (\textbf{B}) Lorenz system and the (\textbf{C}) actuated turbulent
    boundary layer, respectively.
\label{Fig:CNMc}
  }
\end{figure}

To disambiguate the effect of internal dynamics from actuation or external input,
we generalize CNM to include control $\bm{b}$.
The transition probabilities $\bm{Q}(\bm{b})$ and transition times
$\bm{T}(\bm{b})$ are first identified for each actuation setting $\bm{b}$ individually.
The three-step procedure for the propagation of a new control command $\hat{\bm{b}}$
depicted in Fig. \ref{Fig:CNMc}A is then performed. At each iteration, (1) a search for the nearest centroids
from the two closest actuation test cases is performed. (2) Their transition
properties are then identified and (3) averaged to determine the transition of the
state $\bm{\hat{x}}$.
More details of the CNMc algorithm are provided in Appendix \ref{Sec:CNMc}.
CNMc is applied to two systems at new control conditions, the Lorenz attractor
and the actuated turbulent boundary layer. The Lorenz system with $\rho=28$ is
interpolated from two test cases with $\rho=26$ and $\rho=30$ and the boundary layer
with actuation parameters
$\lambda^+=1000$, $T^+=120$ and $A^+=30$ is interpolated
from cases with 
$\lambda^+=1000$, $T^+=120$, $A^+=20$
and
$\lambda^+=1000$, $T^+=120$, $A^+=40$. The CNMc settings are listed in
Table~\ref{Tab:CNMcSettings}.
Despite the algorithm's simplicity, the main
dynamics are properly captured, as shown by the autocorrelation functions in
Fig.~\ref{Fig:CNMc}B and \ref{Fig:CNMc}C,
and the time series (Fig.~\ref{Fig:CNMc:TS}).
CNMc is cast in the same probabilistic framework as CNM and thereby retains all previously-demonstrated advantages.
As the dynamics are interpolated from centroids that belong to potentially
different trajectories, the resulting motion might be noisier and a larger
number of centroids than regular CNM are typically required.
 
\section{Discussion}

We propose a universal data-driven methodology for modeling nonlinear dynamical systems.
The method builds on prior work in cluster-based Markov modeling and network dynamics.
Cluster-based network modeling has several unique and desirable features.
(1) It is simple and automatable.
Once the various schemes are chosen (e.g., clustering algorithm, transition time, etc),
only two parameters must be selected: the number of clusters $K$ and the Markov chain order $L$.
Too few centroids might oversimplify the dynamics, whereas too many might lead to a
noisy solution.
We note that a high Markov chain order $L$ is not always necessarily advantageous.
Both parameters are problem-dependent and can be automatically optimized.
(2) The method does not require any assumption on the analytical structure of the model,
only some sense of smoothness.
It is always honest to the data.
(3) The offline computational load is low.
In fact, the most expensive step in the process is the occasionally-required snapshot-based proper-orthogonal decomposition (POD) for dimensionality reduction.
After the POD computation, the clustering and network modeling require a tiny fraction of the computational operation.
(4) The recurrence properties are the same as the reference data.
If one cluster is visited multiple times (or seldom) in the data, it will also be a recurrence point of the CNM.
This feature is what enables modeling of problems with rare events.
(5) Long-term integration will never lead to divergence -- unlike, e.g., POD-based models.
The simplicity and robustness, however, have a price. 
On the kinematic side, the simple CNM version cannot extrapolate, e.g., resolve oscillations at higher amplitudes not contained in the data. 
On the dynamic side, we lose the relationship to first principles: The network model is purely inferred from  data, without links to the governing equations. 
In particular, cluster-based models are not natural frameworks for dynamic instabilities, 
as the notion of exponential growth and nonlinear saturation is intimately tied to Galerkin expansions. 
Subsequent generalizations need to overcome these restrictions.
(6) The framework is generalizable allowing control-oriented predictions beyond the training data.
A simple interpolation-based control-oriented extension of CNM is proposed and tested.
Despite its simplicity, CNMc accurately predicts the state dynamics at new operating conditions over the entire sample record.

CNM is found to have a distinct superiority over cluster-based Markov models, 
namely the much longer prediction horizon as evidenced by the autocorrelation function. 
The modeling and prediction capabilities are demonstrated on a number of examples
exhibiting chaos, rare events, and high-dimensionality. 
In all cases, the dynamics are remarkably well represented with CNM;
The temporal evolution of the main flow dynamics, the fluctuation level, the autocorrelation function, and the cluster population are all accurately reproduced.

CNM opens a novel automatable avenue for data-driven nonlinear dynamical modeling and real-time control.
It holds the potential for a myriad of further research directions.
Its probabilistic foundations are naturally extendable to include uncertainty quantification and propagation.
One limiting requirement of CNM is the relatively large statistically-converged training data it requires
compared to other known methods (e.g. ARMA and SINDy).
This requirement could be relaxed through explicit coupling to first-principle equations.
The control-oriented extension may be further refined and more broadly implemented on other applications.

\section*{Acknowledgments}
We are grateful to Themistoklis Sapsis, Steve Brunton, Wolfgang Schröder and Marian Albers
for the stimulating discussions and
for providing some of the employed data. 
\textbf{Funding:} The research was funded by the Deutsche Forschungsgemeinschaft (DFG) in the framework of the research projects SE 2504/2-1.
\textbf{Authors contributions:}
B.R.N. conceptualized the algorithm. B.R.N., R.S. and D.F. performed the
investigation, data analysis and interpretation. R.S. and D.F. wrote the
manuscript and D.F. implemented the software.
\textbf{Competing interests:} The authors declare no competing interests.
\textbf{Data and materials availability:} A CNM Python package along with the data used for this study are
available in the github repository at \href{https://github.com/fernexda/cnm}{github.com/fernexda/cnm}.

\begin{appendices}

\section{Problem settings}
\label{Sec:SM:Results}

Cluster-based network modeling is applied to numerous examples, ranging from analytical systems
to real-life problems using experimental and simulation data. The main
results are summarized in Fig. \ref{Fig:Res:SummaryFig}.
The first two applications are the Lorenz~\cite{Lorenz1963JAS} and Rössler~\cite{Rossler1976PLA} attractors,
typical candidates for dynamical systems analysis.
The following two implementations are one-dimensional systems: electrocardiogram
measurements (ECG)~\cite{Brunton2017NC}, and the dissipative energy from a
Kolmogorov flow~\cite{Farazmand2017SA}.
The last CNM application is a high-dimensional large eddy simulation of an
actuated turbulent boundary layer for skin friction reduction~\cite{Albers2020FTaC}.

In this section, we detail the various systems including the numerical setup and
the CNM modeling parameters. CNM is fully parametrized by the number of clusters $K$ and the model order
$L$. Their selection plays an important role in the model accuracy. The values
used for the various systems are listed in Table \ref{Tab:CNMSettings}. The procedure to
select $K$ and $L$ is detailed in Appendix \ref{Sec:Validation}. The last
column in Table \ref{Tab:CNMSettings} lists the normalized time delays
$t_L/T_0$, where $T_0$ is the fundamental period
computed from the dominant frequency identified from the autocorrelation
function. For purely random signals with no deterministic component, such as the
dissipative energy of the Kolmogorov flow, no characteristic period can be
defined.
\begin{table}[h!]
  \centering
  \caption{
    \textbf{CNM settings for all applications.}
    The number of clusters $K$ and the model order $L$ are listed for the five
    systems.
    The last column $t_L/T_0$ designates the normalized time delay
    corresponding to the selected model order $L$. The fundamental period $T_0$ is computed from the
    dominant frequency of the system, when possible.
  }
  \begin{tabular}{>{\centering}m{2cm}|>{\centering}m{2.5cm}|>{\centering}m{2cm}|>{\centering\arraybackslash}m{3cm}}
    \textbf{System} &  \textbf{Number of clusters} $\bm{K}$ & \textbf{Model order} $\bm{L}$ & $\bm{t_{L}/T_{0}}$\\
    \toprule
    Lorenz & 50 & 22 & 1.7 \\
    \hline
    Rössler & 100 & 2& 0.6\\
    \hline
    ECG & 50 & 23& 0.14 \\
    \hline
    Kolmogorov flow & 200 & 25 & - \\
    \hline
    Boundary layer & 50 & 3 & 0.25\\
  \end{tabular}
  \label{Tab:CNMSettings}
\end{table}
As indicated by the table, the CNM parameters are strongly dependent on the nature
of the systems dynamics.
Physical interpretation of the chosen parameters is provided for each system
in the following.

\subsubsection*{Lorenz system}

The Lorenz system~\cite{Lorenz1963JAS} is a typical candidate for dynamical
system analysis. Despite its low dimension, it exhibits a chaotic behavior.
The motion is characterized by periodic oscillations of growing amplitude in the
'ears' and a random switching between them.
The Lorenz system is driven by a set of three coupled nonlinear ordinary
differential equations (ODEs) given by
\begin{equation}
  \begin{aligned}
    \frac{\mathrm{d} x}{\mathrm{d} t} &=\sigma (y-x) \\
    \frac{\mathrm{d} y}{\mathrm{d} t} &=x (\rho-z) -y \\
    \frac{\mathrm{d} z}{\mathrm{d} t} &=x y-b z\,.
    \label{Eqn:LorenzSystem}
  \end{aligned}
\end{equation}
The selected parameters are $\sigma=10$, $\rho=28$, and $\beta=8/3$
with initial conditions $(-3,0,31)$. The simulation is performed with a
time step $\Delta t=0.015$ for a total of 57\,000 samples.
The numerical integration is performed with the explicit Runge-Kutta method of
5\textsuperscript{th} order using the \texttt{scipy} library from the \texttt{python} programming
language~\cite{Rossum2011Book,Pauli2020Nature}.

The relatively high number of clusters ($K=50$) ensures that each wing is resolved
by two orbits of centroids (see the phase space clustering in Fig.
\ref{Fig:Res:SummaryFig}), and allows to reproduce some of the increasing
oscillation amplitude. $K$ can be increased (decreased) to resolve more (less)
orbits in each ear.
Due to the dynamics complexity and especially the random ear flipping, the
Lorenz system requires a large time delay $t_L$ equivalent to 1.7 rotation.
With lower $L$ values, the trajectory that reaches the ears intersection becomes more
likely to wrongly switch sides.

\subsubsection*{Rössler system}

The Rössler is a three-dimensional system governed by non-linear ordinary differential
equations~\cite{Rossler1976PLA} that read
\begin{equation}
  \begin{aligned}
    \frac{\mathrm{d} x}{\mathrm{d} t} &= -y-z\\
    \frac{\mathrm{d} y}{\mathrm{d} t} &= x+ay\\
    \frac{\mathrm{d} z}{\mathrm{d} t} &= b+z(x-c)\,.
    \label{Eq:RosslerSystem}
  \end{aligned}
\end{equation}
where the parameters are $a=0.1$, $b=0.1$, and $c=14$.
The initial conditions are set to $(1,1,1)$ and the simulation is performed with a
time step $\Delta t=0.01$ for a total of 50\,000 samples.
The Rössler data is also created with the \texttt{scipy} library using the explicit Runge-Kutta method of
5\textsuperscript{th} order.
Similar to the Lorenz system, the Rössler is widely used for dynamical system
analysis. The system also yields chaotic behavior under specific parameters combinations. The
motion is characterized by rotations of slowly growing amplitude in the $x$-$y$ plane,
and intermittent peaks in the $z$ direction.

The Rössler system requires a large number of clusters to ensure
a sufficient centroid coverage in the peak for an accurate reproduction of
this intermittent and fast event.
However, since the trajectory itself is relatively simple, a time-delay $t_L$ of
approximately half of the characteristic period is sufficient ($t_L/T_0=0.6$).

\subsubsection*{Electrocardiogram signal}

An electrocardiogram (ECG) measures the heart activity over time.
Electrodes are placed on the person's skin to deliver an univariate voltage of the cardiac
muscle movements. The time series exhibit the typical pulse associated with the heart beat.
The ECG signal used in this study is from the PhysioNet database~\cite{Goldberger2000C}. The signal time
range is 180 seconds and the sampling frequency is 250~Hz.

Similarly to the Rössler, the ECG requires a large number of clusters $K$ in order to resolve the
quasi-circular phase space trajectory corresponding to the fast heartbeat pulse.
Again, due to the very regular and repetitive nature of the heart activity, a
small time delay $t_L$ is sufficient.

\subsubsection*{Kolmogorov flow}

The Kolmogorov flow is a two-dimensional generic flow defined on a square
domain $\bm{q}=(x,y)$ with $0\leq x\leq L$ and $0\leq y\leq L$, subject to a horizontal sinusoidal forcing
$\bm{f}$, defined by
\begin{equation}
  \label{}
  \bm{f}(x)=\sin(a\>y)\bm{e}_1\,,
\end{equation}
where $\bm{e}_1=(1,0)^T$ is a unit vector in the $x$ direction.
The Kolmogorov flow is a test-bed for various fluid mechanics and turbulence studies~
\cite{Fylliaditakis2018JAMP}.
The temporal evolution of the flow energy $E$, the dissipative energy $D$ and
input energy $I$ are defined by
\begin{align}
  \label{}
  E(t) & = \frac{1}{2L^2}\iint\lvert\bm{u}(\bm{q},t)\rvert^2\mathrm{d}\bm{q} \\
  D(t) & = \nu\frac{1}{L^2}\iint\lvert\bm{\omega}(\bm{q},t)\rvert^2\mathrm{d}\bm{q} \\
  I(t) & = \frac{1}{L^2}\iint\lvert\bm{u}(\bm{q},t)\cdot\bm{f}(\bm{q},t)\rvert^2\mathrm{d}\bm{q}
\end{align}
where $\nu$ is the fluid viscosity and $\omega$ is the vorticity.
The rate of change of the energy is equal to the input energy minus the
dissipation energy, as $\dot{E}=I-D$.
With increasing forcing wave number $a$, the dissipation energy yields
intermittent and random bursts. This behavior makes the dissipation energy a good candidate for
rare events modeling.
The current data were created and generously shared by Farazmand et al.~\cite{Farazmand2016PRE}, with a wavenumber $a=4$ and a
Reynolds number $Re=40$. The total time range is 100\,000 dimensionless time
units with a sampling frequency of 10.

The trajectory in the phase space
spanned by $D$ and its temporal derivative $\dot{D}$ (Fig.
\ref{Fig:Res:SummaryFig}) is particularly complex. The region with higher
clusters density in the left region of the phase space corresponds to the random
fluctuations, and the region with sparser centroids distribution
describes the intermittent energy bursts.
Due to its stochastic nature and the absence of deterministic patterns, the Kolmogorov flow dissipation
energy has been particularly challenging to model.
Remarkably, with sufficiently large $K$ and $L$, CNM is capable of
modeling $D$ with high accuracy.

\subsubsection*{Actuated turbulent boundary layer}

The reduction of viscous drag is crucial for many flow-related application such
as airplanes and pipelines, as it is a major contributor to the total drag.
Many passive~\cite{Bechert1985AIAA,Luhar2016JoT} and
active~\cite{Du200Science,Quadrio2011PTRSA} actuation techniques have been
investigated to reduce the skin-friction drag.
In this study, skin-friction reduction on a turbulent boundary layer is achieved
by means of a spanwise traveling surface wave
\cite{Albers2020FTaC,Fernex2020PRF}.

The waves are defined by their wavelength $\lambda^+$, period $T^+$ and amplitude
$A^+$. The superscript $^+$ denotes variables scaled with the friction velocity
and the viscosity. Details about the computational setup can be
  found in Albers et al.~\cite{Albers2020FTaC}.
The actuation parameters are
$\lambda^+=1000$, $T^+=120$, $A^+=60$.
The total time range in ${}^+$ units is 846 and the sampling frequency is 0.5,
resulting into 420 snapshots.
The velocity field is given by $\bm{u}(\bm{q},t^+)$, where $\bm{q}=(x^+,y^+,z^+)$
in the Cartesian coordinates with $x^+\in [2309,4619]$, $y^+\in [0,692]$
and $z^+\in[0,1000]$.

Clustering of large high-dimensional datasets is costly.
The required distance computation between two snapshots $\bm{u}^m$ and
$\bm{u}^n$
\begin{equation}
  \label{Eq:Dist}
  d(\bm{u}^m,\bm{u}^n)=\lVert\bm{u}^m-\bm{u}^n\rVert_{\bm{\Omega}}
\end{equation}
is computationally very expensive. Here, the norm is defined as
\begin{equation}
  \label{Eq:Norm}
  \lVert\bm{u}\rVert_{\bm{\Omega}} = \sqrt{(\bm{u},\bm{u})_{\bm{\Omega}}}
\end{equation}
and the inner product in the Hilbert space $\mathcal{L}(\bm{\Omega})$ of
square-integrable vector fields in the domain $\bm{\Omega}$ is given by
\begin{equation}
  \label{Eq:IP}
  (\bm{u},\bm{v})_{\bm{\Omega}} = \int_{\bm{\Omega}}
  \bm{u}(\bm{q})\bm{v}(\bm{q})
  \, \mathrm{d}\bm{q}\,.
\end{equation}
For high-dimensional data such as the boundary layer velocity field, data compression with lossless
proper orthogonal decomposition (POD) can reduce the computational cost of
clustering.
Here, a snapshot $\bm{u}^m$ is exactly expressed by the POD expansions as
\begin{equation}
  \label{}
  \bm{u}(\bm{q},t) = \bm{u}_0(\bm{q}) + \sum\limits_{i=0}^{M-1}
  a_i(t)\bm{\Phi}_i(\bm{q})\,,
\end{equation}
where $\bm{u}_0$ is the mean flow, $\bm{\Phi}_i$ denotes the POD modes, and
$a_{i}(t)$ the corresponding mode coefficients.
As shown by Kaiser et al.~\cite{Kaiser2014JFM}, the distance computation
(\ref{Eq:Dist}) can be alternatively performed with the mode coefficients instead of the
snapshots, as
\begin{align}
d(\bm{u}^m,\bm{u}^n) & = \lVert\bm{u}^m-\bm{u}^n\rVert_{\bm{\Omega}}\label{Eq:DistSnap}\\
  & = \lVert \bm{a}^m - \bm{a}^n\rVert\label{Eq:DistPOD}.
\end{align}
Hence, $\bm{a}^m = [a_1^m,\ldots,a_{M-1}^m]$ becomes the POD representation of snapshot
$m$ at time $t^m=m\Delta t$.
Eq. (\ref{Eq:DistPOD}) is computationally much lighter than (\ref{Eq:DistSnap}).
Despite the additional autocorrelation matrix computation for the POD process,
the data compression procedure remains very beneficial for large numerical grids. According to
~\cite{Li2020JFM}, the computational savings amount to
\begin{equation}
  \label{Eq:Savings}
  \frac{M+1}{2J\times I \times K}\,,
\end{equation}
where $M$ is the number of snapshots, $K$ the number of clusters, $I$ the number of $k$-means inner
iterations, and $J$ is the number of random centroids initializations.
For typical values ($K\sim 10$, $I\sim 10K$,  and $J\sim 100$),
the saving are one or two orders magnitude.
Furthermore, POD is computed only once for each dataset and will benefit all
future clusterings performed on that dataset.

The actuated turbulent boundary layer at the used actuation settings exhibits
synchronization with the actuation wave.
The dynamics show quasi limit-cycle behavior
with superimposed wandering. Therefore, a low
number of centroids are sufficient to capture the dynamics.
If desired, the limit-cycle meandering associated with higher frequency turbulence can be
resolved with a larger set of centroids.
The selected value of $K=50$ is a compromise between a sufficient resolution of
the turbulence scales (64\% of the data fluctuation is resolved) and a reasonable model complexity.
The dynamics are well captured with a low model order
$L$, equivalent to a time-delay of a quarter of the actuation period.

\section{Cluster-based network modeling methodology}
\label{Sec:CNMMethod}

Robust probability-based data-driven dynamical modeling for complex nonlinear systems has
the potential to revolutionize our ability to predict and control these systems. Cluster-based
network models (CNM) reproduces the dynamics on a directed network~\cite{Barabasi2003SA}, where the nodes are
the coarse-grained states of the system. The transition properties between the nodes are based
on high-order direct transition probabilities identified from the data. The model methodology is
applied to a variety of dynamical systems, from canonical problems such as the Lorenz attractor
to rare events to high degrees of freedom systems such as a boundary layer flow simulation.
The general methodology is illustrated in Fig. \ref{Fig:Meth:CNM} with the Lorenz system and is detailed in the
following.

The first step is the data collection, where a set of $M$ states $\bm{x}^m$, $m=1,\ldots,M$, also called observations or
snapshots, are collected from a dynamical system. They are equally spaced in time
by $\Delta t$, so that $\bm{x}^m=\bm{x}(m\Delta t)$. There is no restriction regarding
the type of system nor the state dimension.

The second step is the identification of the network nodes using an unsupervised
clustering algorithm that groups the snapshots into $K$ clusters $\mathcal{C}_k$,
$k=1,\ldots,K$. In this study, we employ the $k$-means++
algorithm~\cite{MacQueen1967Proceeding,Lloyd1982IEEE,David2006} for its
simplicity and ability to compute physically
meaningful and interpretable centroids.
The algorithm performs an iterative search for an optimal centroid distribution
that increases the inner-cluster similarity, by executing the following steps:
\begin{description}
  \item[Step 1:] The initial centroid distribution $\bm{c}_k$ is randomly generated.

  \item[Step 2:] Each snapshot is affiliated to its closest centroid, following the cluster
    affiliation function $k$ defined as
\begin{align}
      k(\bm{x}^m) &= \text{arg } \underset{i}{\text{min}} \lVert\bm{x}^m -
      \bm{c}_i\rVert\,,
    \end{align}
    where $\lVert \bm{x} \rVert=\sqrt{\bm{x}\cdot\bm{x}}$.
The function $k$ maps, for each state $\bm{x}^m$, the index of the closest
    centroid.

  \item[Step 3:] The inner-cluster variance $J$ of this centroid distribution is computed as 
\begin{equation}
      \label{Eq:Var}
      J(\bm{c}_1,\ldots,\bm{c}_K)=\sum\limits_{k=1}^K\sum\limits_{\bm{x}^m\in\mathcal{C}_k} \lVert \bm{x}^m -
      \bm{c}_k\rVert^2\,.
    \end{equation}

  \item[Step 4:] The centroid positions are updated by averaging the
    state snapshots within the corresponding cluster
\begin{equation}
      \bm{c}_k=\frac{1}{n_k}\sum\limits_{\bm{x}^m\in\mathcal{C}_k}\bm{x}^{m}\,,
    \end{equation}
where $n_k$ is the number of snapshots in cluster $\mathcal{C}_k$.
\end{description}
Steps 2 to 4 are repeated until the inner-cluster variance $J$ is minimized below a specified
tolerance.
Let the vector $\bm{\mathcal{K}}=[\mathcal{K}_1,\ldots,\mathcal{K}_I]$,
$\mathcal{K}_i\in [1,K]$ contain the indexes of all consecutively visited
clusters over the entire time sequence, such that $\mathcal{K}_i$ is the index
of the $i$th visited cluster.
The first and last clusters are $\mathcal{C}_{\mathcal{K}_1}$ and
$\mathcal{C}_{\mathcal{K}_I}$, respectively.
The size $I$ of $\mathcal{K}$ is equal to the number of transitions
between $K$ centroids over the entire ensemble plus one.
The
transition time between the sequential clusters is not constant and depends on
the state velocity in the phase space.
The vector $\bm{\mathcal{K}}$ constitutes the starting point to identify the transition
properties between the centroids.

Following the identification of the centroids as the network nodes, the
third step of the CNM algorithm characterizes the motion along
the nodes.
The dynamics are constructed as linear transitions between centroids based on the
transition probabilities and the transition times.
After a centroid is reached, the next destination is identified
using the direct transition probability tensor $\bm{Q}$.
The tensor $\bm{Q}$ ignores inner-cluster residence probability and only considers inter-cluster transitions.
One novelty of this CNM implementation is to model the direct transition
probabilities $\bm{Q}$ using an $L$-order
Markov model, which is defined as a conditional probability $\mathrm{Pr}\left(\mathcal{K}_i\lvert\mathcal{K}_{i-1},\ldots,\mathcal{K}_{i-L}\right)$.
In this context, high-order Markov models are equivalent to time-delay
embedding, which are well known in dynamical systems~\cite{Takens1981}.
The benefit of time-delay coordinates is elaborated in Appendix
\ref{Sec:CNMUpgrade}.
For a second-order model, $\bm{Q}\in\mathcal{R}^{K\times K\times K}$ is a
third-order tensor and the probability to move to $\mathcal{C}_l$, having previously
visited $\mathcal{C}_k$ and $\mathcal{C}_j$, is inferred from the data and given by
\begin{equation}
  \label{Eqn:HOMM}
  Q_{l,k,j}=\mathrm{Pr}(\mathcal{K}_{i}=l\lvert\mathcal{K}_{i-1}=k,\mathcal{K}_{i-2}=j)=\frac{n_{l,k,j}}{n_{k,j}}\,.
\end{equation}
$n_{l,k,j}$ designates the number of transitions to $\mathcal{C}_{l}$, having
previously visited $\mathcal{C}_{j}$ and $\mathcal{C}_{k}$, and $n_{k,j}$ is
the number of transitions departing from $\mathcal{C}_{k}$ coming from $\mathcal{C}_{j}$, regardless
of the destination.
Note that inner cluster iterations are not possible, by the very definition of a
direct transition, such that $Q_{j,j}=\mathrm{Pr}(\mathcal{K}_i=j|\mathcal{K}_{i-1}=j)=0$.

The transition time designates the time required to travel from one centroid to
the next. In this CNM implementation, the transition time is defined as half of
the sum of the residence times in two sequential clusters, as illustrated in
Fig. \ref{Fig:Meth:Time} for a first-order model.
Let $t^n$ and $t^{n+1}$ be the time of the first and last snapshots to enter
and, respectively, leave $\mathcal{C}_k$ at the $n$th iteration. The iterations designate the sequential jumps between the centroids.
The residence time $\tau^n$ in $\mathcal{C}_k$ is
\begin{equation}
\tau^n := t^{n+1} - t^{n}\,.
\end{equation}
Following this definition, the individual transition time $\tau_{k,j}^n$ from
centroid $\bm{c}_{j}$ to
centroid $\bm{c}_{k}$ is given by
\begin{equation}
\tau_{k,j}^n = \frac{t^{n+1} - t^{n-1}}{2} = \frac{\tau^{n-1}+\tau^{n}}{2}\,.
\end{equation}
The transition time $T_{k,j}$ is the average of all transition times
$\tau_{k,j}^n$ between centroids $\bm{c}_{j}$ and $\bm{c}_{k}$ as
\begin{equation}
T_{k,j} = \frac{1}{n_{k,j}}\sum_{n=1}^{n_{k,j}}\tau_{k,j}^n\,.
\end{equation}
Consistent with the direct transition matrix $\bm{Q}$ for an $L$-order chain,
the transition time $\bm{T}$ also depends on the $L-1$ previously-visited
centroids.

For large time delays (hence large $L$), the process could yield to two storage-intensive $L+1$-dimensional tensors
$Q$ and $T$ with $K^{L+1}$ elements.
For instance, clustering with $K=20$ clusters and an order $L=10$, the
tensors would contain $20^{11}$ elements, which exceeds the storage capacities
of most computers.
The expensive tensor creation and storage is circumvented by
a lookup table (LUT), where only non-zero entries that correspond to actual transitions are
retained. Thus, the tensors are replace by a simple array indexing operation.
The look-up tables are typically orders-of-magnitude smaller than the full
tensors.
As illustration, let's consider the example in Fig. \ref{Fig:Meth:CNM} but
with fictional transition properties.
Here, all centroids
have only one possible destination, except $\bm{c}_2$, where the state can
transit to either
$\bm{c}_3$ or $\bm{c}_5$ with an assumed equal probability for simplicity
($Q_{3,2,1}=Q_{5,2,1}=0.5$). The 2\textsuperscript{nd}-order LUT for this example is
illustrated in Table \ref{Tab:Meth:LUT}. 
The transition times are randomly chosen for this fictional example.
\begin{table}[h!]
  \centering
  \caption{
    \textbf{Lookup table (LUT) of the transition properties.}
    The storage-intensive $L+1$-dimensional tensors $\bm{Q}$ and $\bm{T}$ with
    $K^{L+1}$ elements are replaced by a lookup table, where
    only non-zero entries that correspond to actual transitions are
    retained. The storage requirements are orders of magnitude smaller than that
    of the full tensors.
  }
  $
  \begin{array}{l|ccc|cc|}
    \# & l & k & j & Q_{l,k,j} & T_{l,k,j} \\
    \toprule
    1 & 1 & 4 & 3 & 1 &   2 \\ 2 & 1 & 6 & 5 & 1 &   1 \\ 3 & 2 & 1 & 4 & 1 &   0.5 \\ 4 & 2 & 1 & 6 & 1 &   0.25 \\ 5 & 3 & 2 & 1 & 0.5 & 2 \\ 6 & 4 & 3 & 2 & 1 &   4 \\ 7 & 5 & 2 & 1 & 0.5 & 1\\ 8 & 6 & 5 & 2 & 1 &   2\\ \end{array}
  $
  \label{Tab:Meth:LUT}
\end{table}
If the state is in $\bm{c}_2$ ($k=2$) having visited $\bm{c}_1$
before ($j=1$), the next destination is probabilistically chosen
between $Q_{3,2,1}$ ans $Q_{5,2,1}$ (lines 5 and 7), yielding a transition to
either centroid 3 or centroid 5.
If the selected destination is $\bm{c}_5$, the corresponding time to this
transition is read from $T_{5,2,1}$ (entry \# 7).

The fourth and final step of the CNM procedure is state propagation.
We assume a linear motion between the centroids.
The instantaneous state
between two centroids $\bm{c}_{j}$ and $\bm{c}_{k}$ is defined as
\begin{equation}
  \bm{x}(t) = \alpha_{kj} (t) \bm{c}_{k} + [1 - \alpha_{kj}(t)] \bm{c}_{j},\text{
  with } \alpha_{kj}=\frac{t_{j}-t}{T_{k,j}}\,,
\end{equation}
where $t_{j}$ is the time when the centroids $\bm{c}_{j}$ is left and
$T_{k,j}$ is the transition time from $\bm{c}_{j}$ to $\bm{c}_{k}$.
The motion between the centroids may be interpolated with splines for
smoother trajectories.
As CNM is purely data-driven, the model
quality is directly related to that of the training data. More specifically, the sampling frequency
and total time range must be selected, such that all relevant dynamics are captured and are
statistically converged.
 
\section{Validation}
\label{Sec:Validation}

This section presents two metrics used to evaluate the model performance,
namely the autocorrelation function and the cluster probability vector.
We note the occasional need for additional or different metrics to validate
models of certain systems, such as the probability distribution function of a
system with rare events.

\subsubsection*{Autocorrelation function}

The direct comparison of time series of complex systems is often pointless, as the
trajectories might rapidly diverge even when the dynamical features are
preserved.
This is especially true for chaotic systems, where a slight
change in initial conditions leads to completely different
trajectories.
Following Protas et al.~\cite{Protas2015JFM}, the cluster network model
is validated based on the computed and predicted autocorrelation
function of the state vector, defined as
\begin{equation}
  R(\tau)=\frac{1}{T-\tau}\int\limits_{0}^{T-\tau}\left(\bm{x}(t),\bm{x}(t+\tau)\right)\mathrm{d}t,\,\tau\in
  [0,T]\,.
\end{equation}
where $(\,,)$ designates the inner product, defined as
\begin{equation}
  (\bm{x},\bm{y}) = \bm{x}\cdot\bm{y}^T\,.
\end{equation}
This function avoids the problem of comparing two trajectories with finite dynamic prediction
horizons due to phase mismatch. The autocorrelation function also yields the fluctuation energy
at vanishing delay $R(\tau=0)$ and can be used to infer the spectral behavior.

In case of the cluster-based Markov model (CMM) (c.f. Appendix
\ref{Sec:CNMUpgrade}), the time integration quickly leads to the average
state and is not indicative for the range of possible initial conditions. Hence,
$K$ trajectories are considered starting sequentially in each of the $K$
clusters, such that
$\bm{p}^k(t=0)=[\delta_{1k},\ldots,\delta_{Kk}]^{\mathrm{T}}$, where 
$\delta$ is the Dirac delta function.
The autocorrelations are weighted with the cluster probability $p_{k}^\infty$
as,
\begin{equation}
  \hat{R}(\tau)=\sum_{k=1}^{K}p_k^\infty
  \frac{1}{T-\tau}\int\limits_{0}^{T-\tau}\left(\hat{\bm{x}}^k(t),\hat{\bm{x}}^k(t+\tau)\right)\mathrm{d}t,\,\tau\in
  [0,T]\,,
\end{equation}
where $\hat{\bm{x}}^k$ is the CMM-modeled trajectory initialized in centroid
$k$.

\subsubsection*{Cluster probability distribution}

The cluster probability distribution $\bm{p}=[p_1,\ldots,p_K]$,
provides the probability to be in a specific cluster.
It indicates whether the model trajectories populate the phase space similarly
to the reference data.
The cluster probability distribution of the reference data
is computed as
\begin{equation}
  \label{Eq:CPV:Data}
  p_k = \frac{n_k}{M}\,,
\end{equation}
where $n_k$ is the number of snapshots affiliated to cluster $\mathcal{C}_k$ and
$M$ is the total number of snapshots.
In CMM, the state is iteratively propagated with time steps $\Delta t$, so that
$t^l=l\Delta t$. The asymptotic cluster probability distribution for CMM $\bm{p}^\infty$ is determined for
$l\to\infty$ by
\begin{equation}
  \label{Eq:CPV:CMM}
  \bm{p}^\infty = \lim\limits_{l\to\infty}\bm{P}^l\bm{p}^0\,,
\end{equation}
where $\bm{p}_0$ is the initial condition and $\bm{P}$ is the transition matrix
(c.f. Appendix \ref{Sec:CNMUpgrade}).
For CNM, the asymptotic probability
distribution $p_k^\infty$ is obtained for a long-enough  time horizon
$T_0$ as the sum of the residence times $\tau^i$ in $\mathcal{C}_k$ divided
by the total simulation time,
\begin{equation}
  \label{}
  p_k^\infty = \frac{\sum \tau^i}{T_0}\,.
\end{equation}
The cluster probability distributions for the data, CMM, and CNM with varying
order are presented in Fig. \ref{Fig:CROM:CPV} and discussed in Appendix \ref{Sec:CNMUpgrade}.

\section{Parameters selection}
\label{Sec:ParamSelection}

CNM is parametrized by the number of clusters $K$
and the model order $L$. Both parameters are problem-dependent and can be
optimized to achieve the highest prediction accuracy.
This section details the current parameters selection process
 and provides guidelines and recommendations for other datasets.

The number of clusters $K$ determines the resolution level.
A small number of clusters will mostly captures the dominant behavior at the
macro level, which is suitable for simple dynamics such as a limit cycle.
Few centroids also enable  easier interpretation of the results and physical
insights.
Higher $K$ allows to accurately model more complex systems, possibly at the
micro scale level, and to uncover a broader range of dynamics including transitional events  and higher
frequency components.
Excessively large $K$ values could be however detrimental, as adjacent trajectories get clustered separately despite
describing nearly the same motion. Furthermore, the trajectory of an over-clustered system
might introduce an artificial low-amplitude and high frequency noise component resulting from
sequential small jumps between misaligned centroids.

\begin{figure}[h!]
  \begin{center}
    \includegraphics[width=7cm]{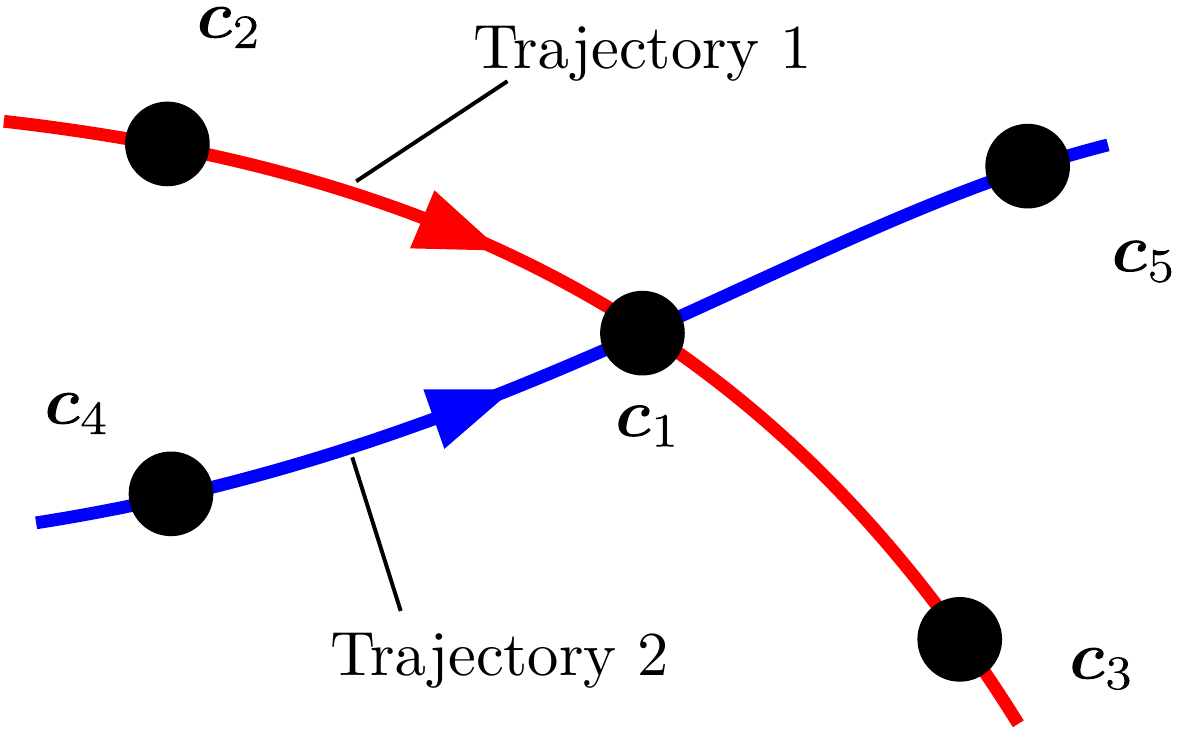}
  \end{center}
  \caption{
    \textbf{Improved accuracy with higher model order $L$.}
    In this example, trajectory 1 and 2 intersect at
    $\bm{c}_1$.
    With $L=1$, the state can possibly leave the trajectory it is
    following, \textit{e.g.} trajectory 1, and wrongly transit to trajectory 2
    after leaving centroids $\bm{c}_1$.
    A 2\textsuperscript{nd}-order model ensure that the state remains on the correct
    trajectory.
  }
  \label{Fig:TrajCrossing}
\end{figure}

The model order $L$ is synonymous with time-delay embeddings in units of
past centroids.
Increasing $L$ allows modeling of more realistic trajectories, especially when the
dynamics have centroids located at the intersections of multiple trajectories,
as illustrated in Fig. \ref{Fig:TrajCrossing}.
In this example, trajectory 1 propagates sequentially through centroids $2\to 1\to
3$ and trajectory 2 through centroids $4\to 1 \to 5$. With a first-order model,
the state at $\bm{c}_1$ can possibly transit to $\bm{c}_3$ or to $\bm{c}_5$,
regardless of the previous path. Such a low-order model is thus associated with
an increased risk of selecting the wrong trajectory.
With a second-order model, however, the previous centroid is taken into account
in the conditional probability and the
transition from $\bm{c}_1$ remains on the correct trajectory,
ensuring a more accurate motion.
Complex dynamics with multiple intersected trajectories require larger order
$L$.
For better interpretability, $L$ can be converted into the
approximate corresponding time delay $t_L$, defined as
\begin{equation}
  \label{Eq:tL}
  t_L = L\overline{T}\,,
\end{equation}
where $\overline{T}$ is the overall average transition time.

\begin{figure}[]
  \begin{center}
    \includegraphics[width=16cm]{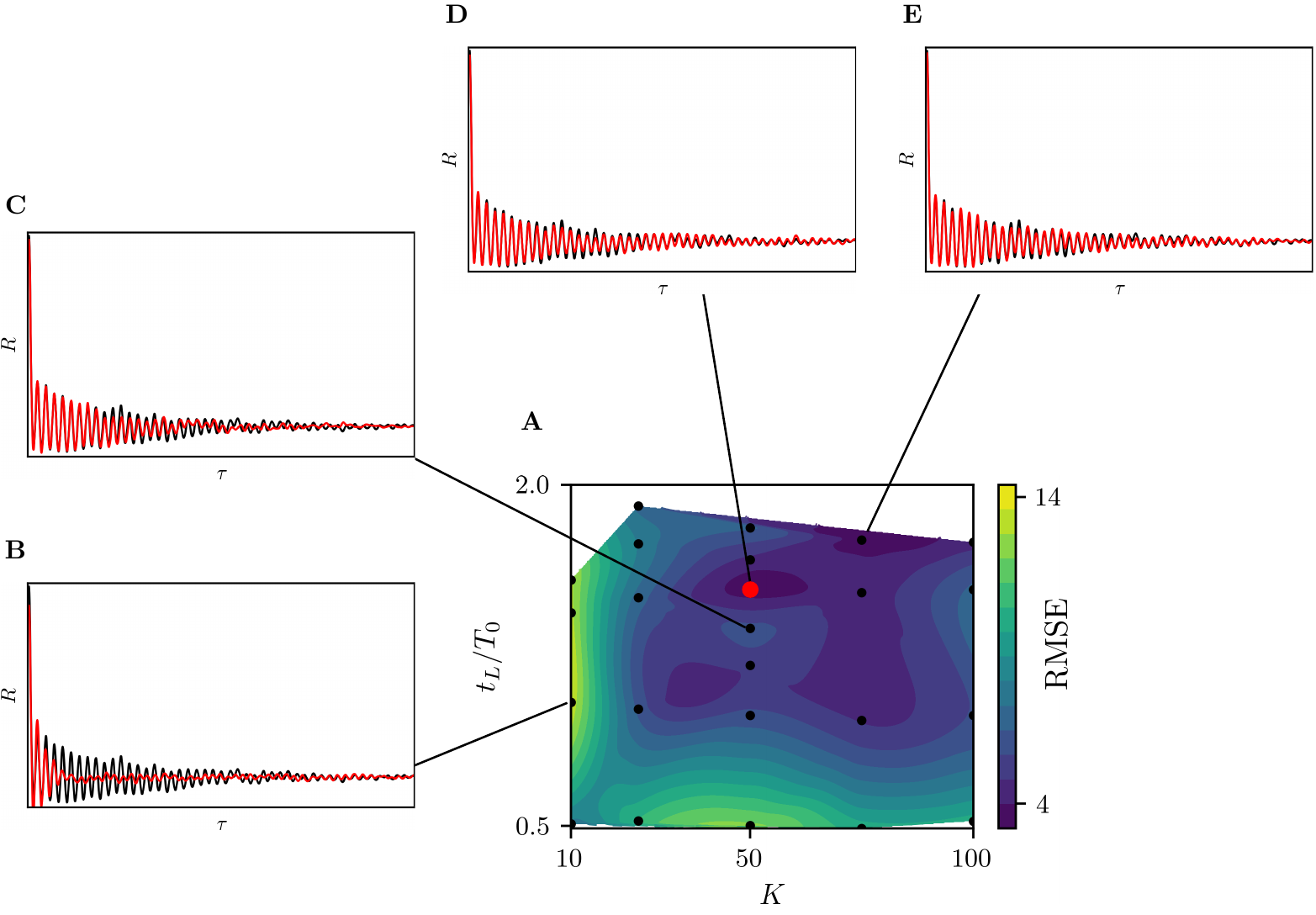}
  \end{center}
  \caption{
    \textbf{Selection of the number of centroids $\bm{K}$ and model order $\bm{L}$.}
(\textbf{A}) Root-mean square error distribution of the autocorrelation
    function $R(\tau)$ of the reference data and that of the model for the
    Lorenz system.
The results are shown for a range of $K\in[10,100]$ and normalized time delays
    $t_L/T_0\in[0.5,1.8]$, where $T_0$ is the fundamental period of the system.
The winning configuration is indicated with a red dot.
(\textbf{B}) to (\textbf{E}) Autocorrelation function of the data (black) and CNM (red)
    for varying $K$ and $L$.
A small number of centroids $K=10$ produces poor dynamics (\textbf{B}). The
    distribution presents local minima and maxima (\textbf{C}). The agreement
    between the reference and modeled $R(\tau)$ increases for larger $K$ and $L$
    values ((\textbf{D}) and (\textbf{E})).
  }
  \label{Fig:HeatMapRMSE}
\end{figure}

The values of $K$ and $T$ are problem-dependent.
In this work, $K$ and $L$ are selected from a parametric study that minimizes
the root mean square error (RMSE) of the autocorrelation function $R(\tau)$ of
the reference data and that of the model (see Appendix \ref{Sec:Validation} for details about
$R(\tau)$), defined as
\begin{equation}
  \label{Eq:RMS-Rxx}
  \text{RMSE} = \sqrt{\frac{1}{N_R}\sum\limits_{n=1}^{N_R}\left(R_n-\hat{R}_n\right)^2}\,,
\end{equation}
where $\hat{R}$ is the modelled autocorrelation and $N_R$ is the maximum lag number.
The RMSE distribution for the Lorenz system
for varying $K$ and $L$ is presented in Fig. \ref{Fig:HeatMapRMSE}A.
The results are shown for a range of $K\in[10,100]$, and normalized time delay
$t_L/T_0\in[0.5,1.8]$, where $T_0$ is the fundamental period of the system.
The black dots denote the computed configurations.
The RMSE does not change linearly with the number of cluster and the
model order.
The error distribution exhibits regions of high and low error.
A low $K$ and low $L$ configuration produces poor dynamics (Fig. \ref{Fig:HeatMapRMSE}B), which is expected for a complex system like the
Lorenz.  As $K$ and $L$  increase,
the error generally decreases, despite local maxima (Fig.
\ref{Fig:HeatMapRMSE}C). The configuration indicated with the red
dot (Fig. \ref{Fig:HeatMapRMSE}D) is selected for its high accuracy,
which is comparable to other more complex models (e.g., Fig.
\ref{Fig:HeatMapRMSE}E).
This winning configuration consists of $K=50$ clusters and an order $L=22$, that
corresponds to $t_L/T_0=1.7$.

 \section{Comparison between CNM and CMM}
\label{Sec:CNMUpgrade}

The CNM implementation presented in this study builds on two prior cluster-based
modeling methods.
The first, labelled cluster-based Markov model (CMM), propagates the state using a Markov
chain with a constant time step~\cite{Kaiser2014JFM}. The second is
the initial CNM implementation, which introduced realistic transition times between
clusters that yielded a more accurate dynamics~\cite{Fernex2019PAMM,Li2020JFM}.
The present work extends CNM to high-order chains for both the transition
probability and the transition time. The resulting drastic improvements over
the two preceding methods are demonstrated in this section using the Lorenz
system as example.

The starting point for CMM and both CNM variants is the clustering of the
snapshots into $K$ clusters $\mathcal{C}_k$ (for details on the clustering
algorithm, see Appendix \ref{Sec:CNMMethod}).
Clustering reduces and coarse-grains the original potentially high-dimensional data into a set of
centroids $\bm{c}_k$. Both methodologies are described in the main manuscript and are
briefly summarized in the following.
In CMM, the state variable is $\bm{p}=[p_1,\ldots,p_K]$, where $p_k$ is the probability of being in cluster
$\mathcal{C}_k$. The transition from centroid $\bm{c}_{j}$ to $\bm{c}_{i}$ is guided by the
probability transition matrix $\bm{P}=(P_{ij})\in \mathcal{R}^{K\times K}$.
The propagation of $\bm{p}$ in time is performed iteratively in steps of $\Delta t$
and is given at time $t^l=l\Delta t$ by
\begin{equation}
\bm{p}^{l+1} = \bm{P}\bm{p}^{l}\,.
\end{equation}
The state $\bm{x}$ at time $t^l$ is given by
\begin{equation}
  \label{Eq:CROM}
  \bm{x}(t^l) = \sum\limits_{k=1}^K p_k(t^l) \bm{c}_k\,.
\end{equation}
Despite its ability to provide insights into the guiding mechanisms of various
physical systems~\cite{Cao2014EF,Ishar2019JFM}, CMM fails at modeling the
dynamics.
The state vector of cluster probabilities ultimately and unavoidably
diffuses to a fixed point representing the post-transient attractor.

The initial CNM version addresses this issue by introducing realistic transition
times. At each iteration, the state shifts to a subsequent cluster in a
data-inferred time.
The direct transition probabilities $\bm{Q}$ and transition times $\bm{T}$ are
$K\times K$ zero-diagonal matrices, which result from the state
switching centroids at each iteration.
The data-inferred transition times drastically increase accuracy compared to CMM, especially for
periodic dynamics. This algorithm is, however, not well-suited for complex
phase-space trajectories and long time-horizon predictions.
The present CNM variant extends the algorithm to high-order Markov models, where
both past and current states are jointly considered to determine the next
destination cluster and the corresponding transition time.  The optimal order is
problem-dependent and can be optimized, as detailed in Appendix
\ref{Sec:ParamSelection}.
With this latest upgrade, CNM is now capable of modeling any complex nonlinear
dynamical system.

\begin{figure}[h!]
  \begin{center}
    \includegraphics[width=12cm]{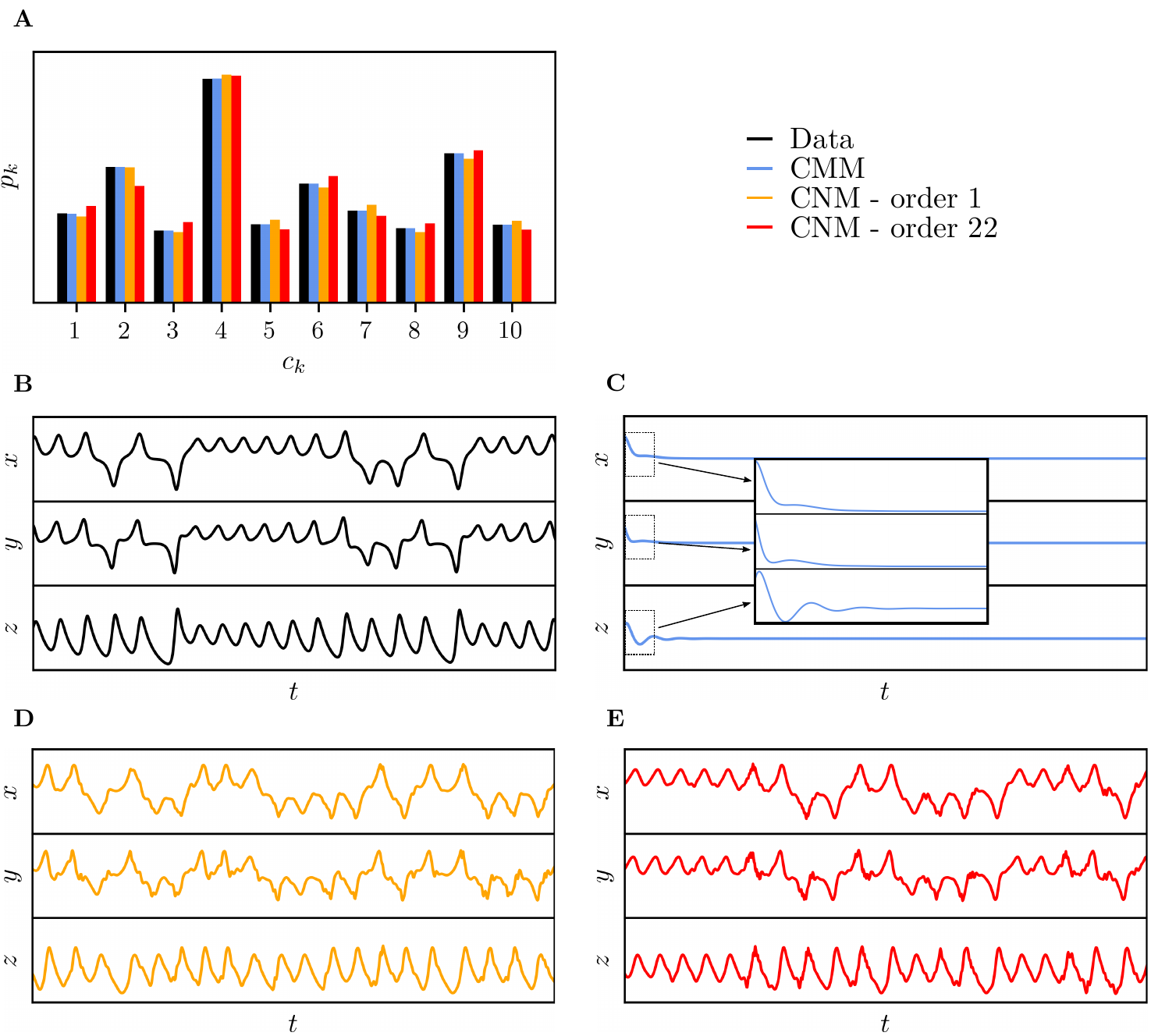}
  \end{center}
  \caption{
    \textbf{Comparison of the cluster probability distributions and time series.}
    (\textbf{A}) Cluster probability distribution of the data, CMM,
    1\textsuperscript{st}-, and 22\textsuperscript{nd}-order CNM, respectively.
    The distributions are shown for 10 centroids for clarity.
    CMM reproduces exactly the probabilities of the reference data, whereas the
    CNM cluster probability distributions agree very well with the data.
    (\textbf{B})-(\textbf{E})Time series of  the data, CMM, 
    1\textsuperscript{st}-, and 22\textsuperscript{nd}-oder CNM, respectively.
    (\textbf{C}) The CMM model fails at predicting any dynamics.
    (\textbf{D}) and (\textbf{E}) Both CNM variants capture
    the oscillations and ear switching of the reference data.
  }
  \label{Fig:CROM:CPV}
\end{figure}

The three methods are benchmarked on the Lorenz system, that is described in
Appendix \ref{Sec:SM:Results}.
Clustering is performed with $K=50$ clusters and the model order is set to $L=22$.
The cluster probability distribution, which indicates whether the model
trajectories populate the phase-space similarly to the data,
is depicted for the
three methods in Fig. \ref{Fig:CROM:CPV}A.
For clarity of presentation and interpretation, the results are shown for only 10
centroids instead of 50. The coarsening is performed by affiliating the
CNM-generated snapshots to the closest of the 10 centroids.
The converged probability distribution for CMM matches exactly that of
the data, as already reported in the literature~
\cite{Kaiser2014JFM,Osth2015JWEIA}, whereas those for
both CNM variants yield a very good agreement.

The time series of the reference data and of the three models are shown in Fig.
\ref{Fig:CROM:CPV}B to \ref{Fig:CROM:CPV}E. The reference time series
(\ref{Fig:CROM:CPV}B) depict the growing amplitude
oscillations in both ears as well as the random ear switching.
The CMM temporal evolution in \ref{Fig:CROM:CPV}C quickly asymptotes toward a fix value after a few
oscillations, thus demonstrating the model inability to resolve any meaningful
dynamics.
Conversely, both CNM models appear to properly duplicate the reference time
series (Fig. \ref{Fig:CROM:CPV}D and Fig. \ref{Fig:CROM:CPV}E).

\begin{figure}[h!]
  \begin{center}
    \includegraphics[width=15cm]{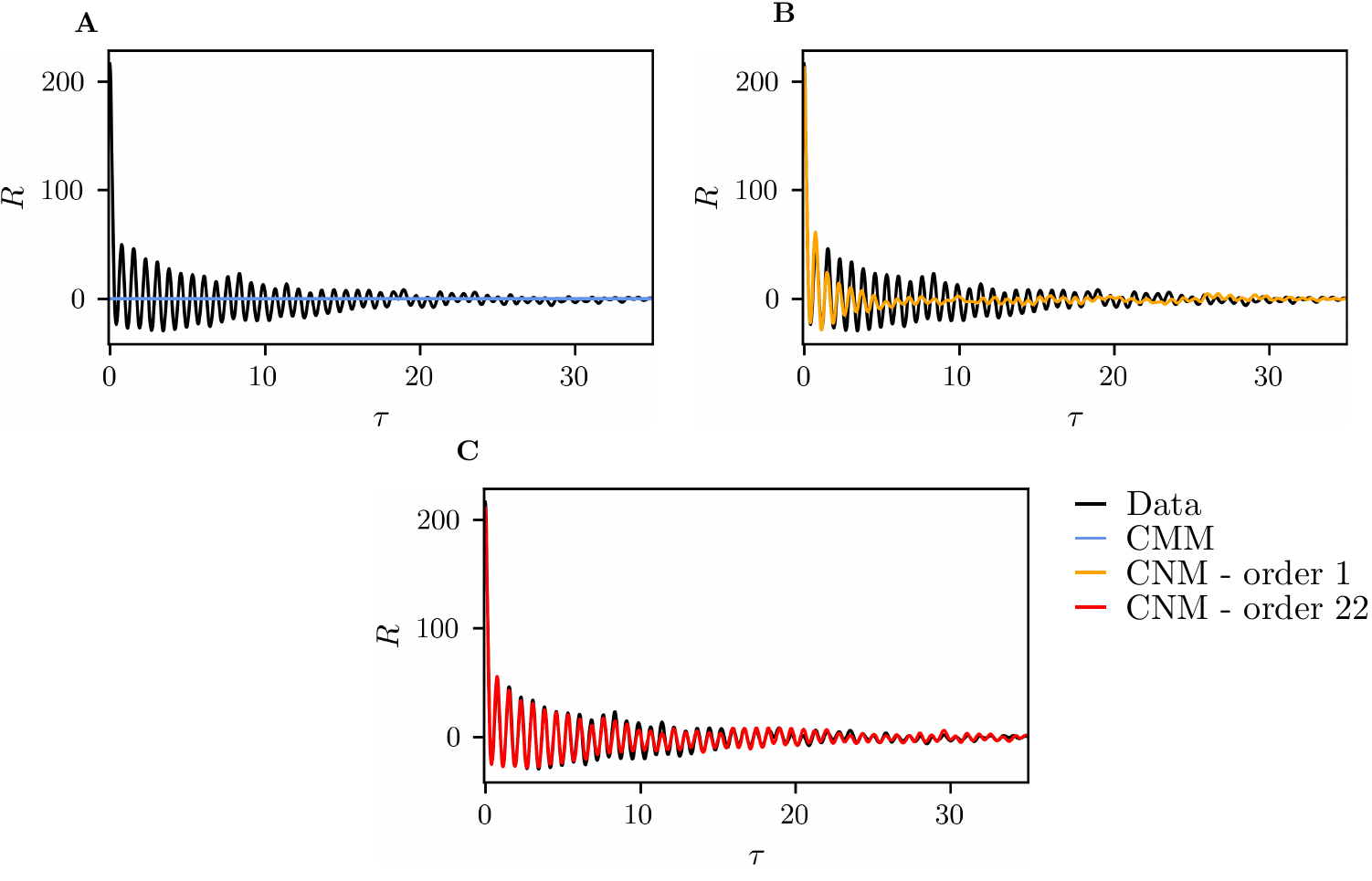}
  \end{center}
  \caption{
    \textbf{Comparison of the autocorrelation function.}
    Autocorrelation function of (\textbf{A}) the data, (\textbf{B}) CMM, (\textbf{C})
    1\textsuperscript{st}-, and (\textbf{D}) 22\textsuperscript{nd}-order CNM, respectively.
    The superiority of the high-order CNM over the two other methods is clearly
    visible.
    (\textbf{A}) CMM yields a flat autocorrelation function, demonstrating that no
    dynamics are resolved.
    (\textbf{B}) After a few oscillations, the first-order
    CNM prediction rapidly deteriorates.
    (\textbf{C}) The agreement of the 22\textsuperscript{nd}-order model with the data is
    excellent over the entire time range and confirms the model long
    time-horizon prediction capabilities.
  }
  \label{Fig:CROM:Rxx}
\end{figure}

The benefits of the high-order CNM become apparent when comparing the autocorrelation
function distributions in Fig. \ref{Fig:CROM:Rxx}. As expected, the autocorrelation
function from CMM rapidly drops to zero (Fig. \ref{Fig:CROM:Rxx}A).
The first-order CNM roughly reproduces
the first few iterations, but both amplitude and phase quickly diverge from
those of the data (Fig. \ref{Fig:CROM:Rxx}B).
In contrast, the high-order CNM, accurately duplicates the
oscillations amplitude and phase over the entire range and evidentiates the
correct modeling of the dynamics (Fig. \ref{Fig:CROM:Rxx}C).

\section{Control-oriented cluster-based network modeling}
\label{Sec:CNMc}

\begin{figure}[h!]
  \begin{center}
    \includegraphics[width=15cm]{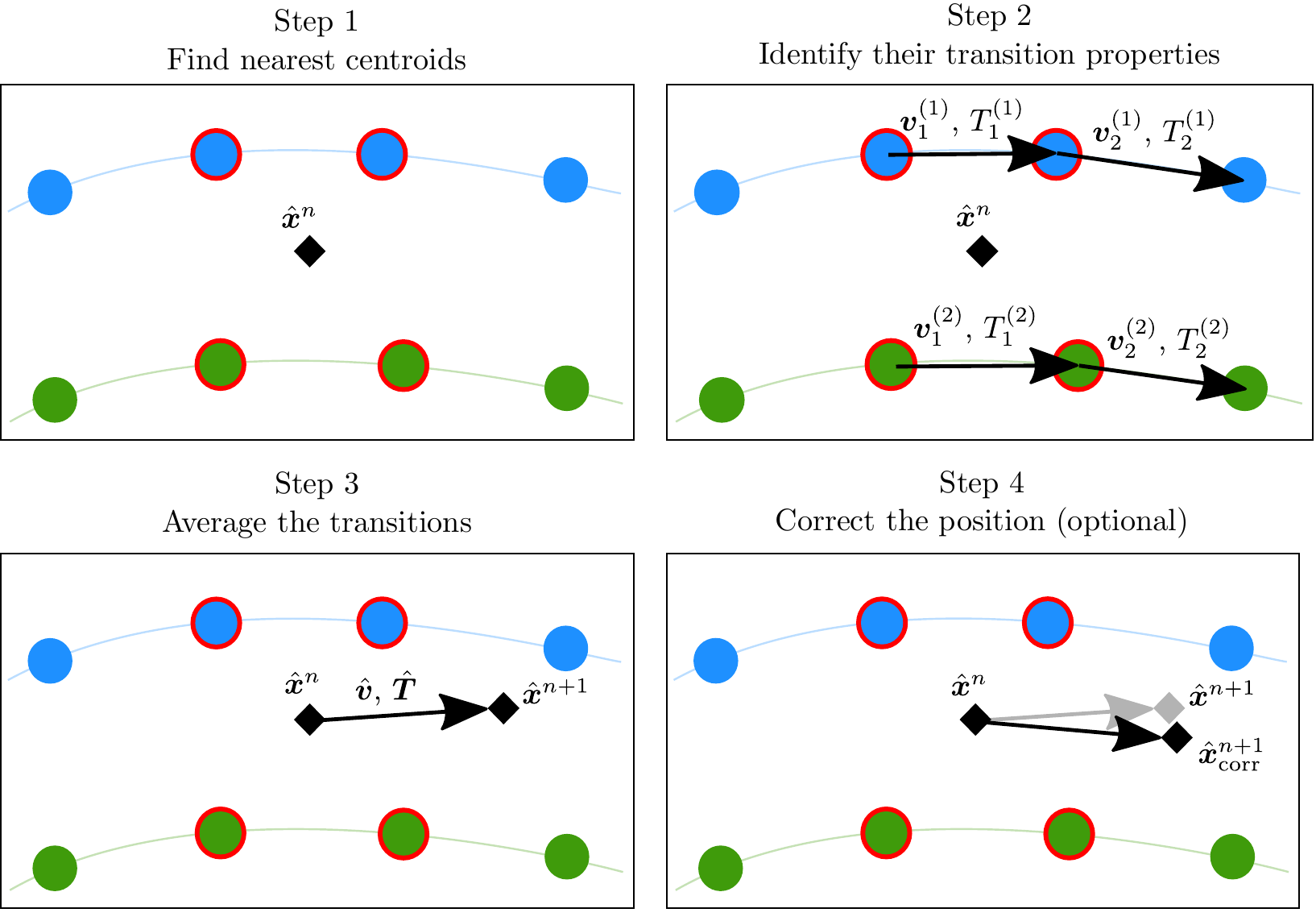}
  \end{center}
  \caption{
    \textbf{Main steps of CNMc.}
    The centroids of two neighboring test cases are identified.
    The state $\bm{x}$ is propagated following 4 steps.
Step 1: The nearest centroids $\bm{c}_{n_k}^{(i),n}$ to the current state $\hat{\bm{x}}^n$ are
    determined (red outer circles).
    Step 2: The transition vector $\bm{v}_{n_k}^{(i)}$ and transition time
    $\bm{T}_{n_k}^{(i)}$ of each nearest centroid are identified.
    Step 3: The transition to the new position $\hat{\bm{x}}^{n+1}$ is
    set by $\hat{\bm{v}}$ and $\hat{T}$, which are the average of the
    nearest neighbors transition vectors and transition times, respectively.
    Step 4: For some systems, an additional position correction is needed to
    avoid a drift of the trajectory toward one of the two attractors.
    These steps are repeated until the total simulation time is reached.
  }
  \label{Fig:Control}
\end{figure}

To disambiguate the effect of internal dynamics from actuation or external input,
we generalize CNM to include control $\bm{b}$.
This enables predictions beyond the training data for new control parameters $\hat{\bm{b}}$.
The main steps of CNMc are illustrated in Fig. \ref{Fig:Control} and the
algorithm is detailed in Algorithm \ref{Algo:CNMc}.

Before we detail the algorithm, we introduce some the relevant parameters and
definitions. Let $I$ be the number of test cases with $I$ different
control terms, where
the superscript $(i)$ designate the $i$\textsuperscript{th} test case,
$i=1,\ldots,I$.
We denote the control term for the new to-be-predicted test case as $\hat{\bm{b}}$.
The distance between
two control terms $\bm{b}^i$ and $\bm{b}^j$ is defined by
\begin{equation}
  \label{Eq:Distb}
  d\left(\bm{b}^{i},\bm{b}^{j}\right)=\lVert \bm{b}^{i}-\bm{b}^{j} \rVert\,,
\end{equation}
where $\lVert .\rVert$ designates the Euclidean norm.
Similarly to all machine-learning methods, CNMc is inherently limited in its
ability to extrapolate beyond the data range. Specifically,
the control parameter $\hat{\bm{b}}$ must lie within the range of available
control space.
Since the method relies on interpolation, two neighboring test cases must be
identified. Using the Euclidean distance (\ref{Eq:Distb}),
the two closest operating conditions with the shortest distance to
$\hat{\bm{b}}$ are determined. Henceforth,
the superscript $(i)$ designates specifically these two
neighboring operating conditions, such that $i=1,2$.
The snapshots of these two operating conditions are separately grouped into $K$ clusters
$\mathcal{C}_k^{(i)}$, $k=1,\ldots,K$. Following CNM for individual operating
conditions (see Appendix \ref{Sec:CNMMethod}), the
transition probabilities $\bm{Q}^{(i)}$ and transition times $\bm{T}^{(i)}$, of
each operating condition are separately computed.
For a realistic initialisation of the new to-be-predicted system, the initial state
$\hat{\bm{x}}_0$ is computed as the average of a centroid from test case ($i=1$)
and its nearest neighbor from test case ($i=2$).

Now that the two neighboring test cases are identified, the state of the new
test case $\hat{\bm{b}}$ can be propagated.
The motion propagation is performed iteratively, following the
four steps illustrated in Fig. \ref{Fig:Control} and detailed in the following.
\begin{description}
  \item[Step 1:] For each neighboring operating condition $(i)$, the $N_k$
    closest centroids to the state $\hat{\bm{x}}^n$ at
    iteration $n$ are identified $\bm{c}_{n_k}^{(i),n}$,
    $n_k=1,\ldots,N_k$. The search is performed using a $k$-d tree, that organizes the data in a tree-like structure to
    rapidly find nearest neighbors~\cite{Bentley1975ACM}.
For low-dimensional data, the $k$-d tree distance metric is typically the
    Euclidean distance.
For the high-dimensional boundary layer data, where the flow is driven by
    large scale actuation, an alternative norm that reduces high-frequency
    low-energy small scale contribution is recommended. One possible norm is the
    $L_{10}$ norm, defined as
    $\lVert \bm{x} \rVert_{10}=\left(\sum_{i=1}^{n}\lvert
    x_i\rvert^{10}\right)^{1/10}$, which favors the
    mode coefficients with large magnitude.

  \item[Step 2:] The transition properties of the $N_k$ nearest centroids are
    determined.
The appropriate $L-1$ past centroids of each neighboring centroid
    $\bm{c}_{n_k}^{(i),n}$ are identified by choosing from all possible
    trajectories leading to $\bm{c}_{n_k}^{(i),n}$ the most similar to the CNMc trajectory, as
    illustrated in Fig. \ref{Fig:PastTraj} for a third-order model.
In this example, the past centroids of $\bm{c}_{n_k}^{(i),n}$ along trajectory 2
    are selected, since trajectory 2 is most aligned with $\hat{\bm{x}}$ up to
    this point.
The next centroid $\bm{c}_{n_k}^{(i),n+1}$ and the transition time are identified from
    the corresponding $\bm{Q}$ and $\bm{T}$.
The motion direction is based on transition
    vectors $\bm{v}_{n_k}^{(i)}$ that span the transition from $\bm{c}_{n_k}^{(i),n}$ to
    $\bm{c}_{n_k}^{(i),n+1}$ as
\begin{equation}
\bm{v}_{n_k}^{(i)}=\bm{c}_{n_k}^{(i),n+1}-\bm{c}_{n_k}^{(i),n}\,.
    \end{equation}
For the first $L-1$ iterations, the CNMc trajectory is not long enough
    to identify the $L-1$ past centroids of the neighbors. In that case, the
    first $L-1$ transition properties are identified from a first-order CNM model.
\begin{figure}[h!]
      \begin{center}
        \includegraphics[width=10cm]{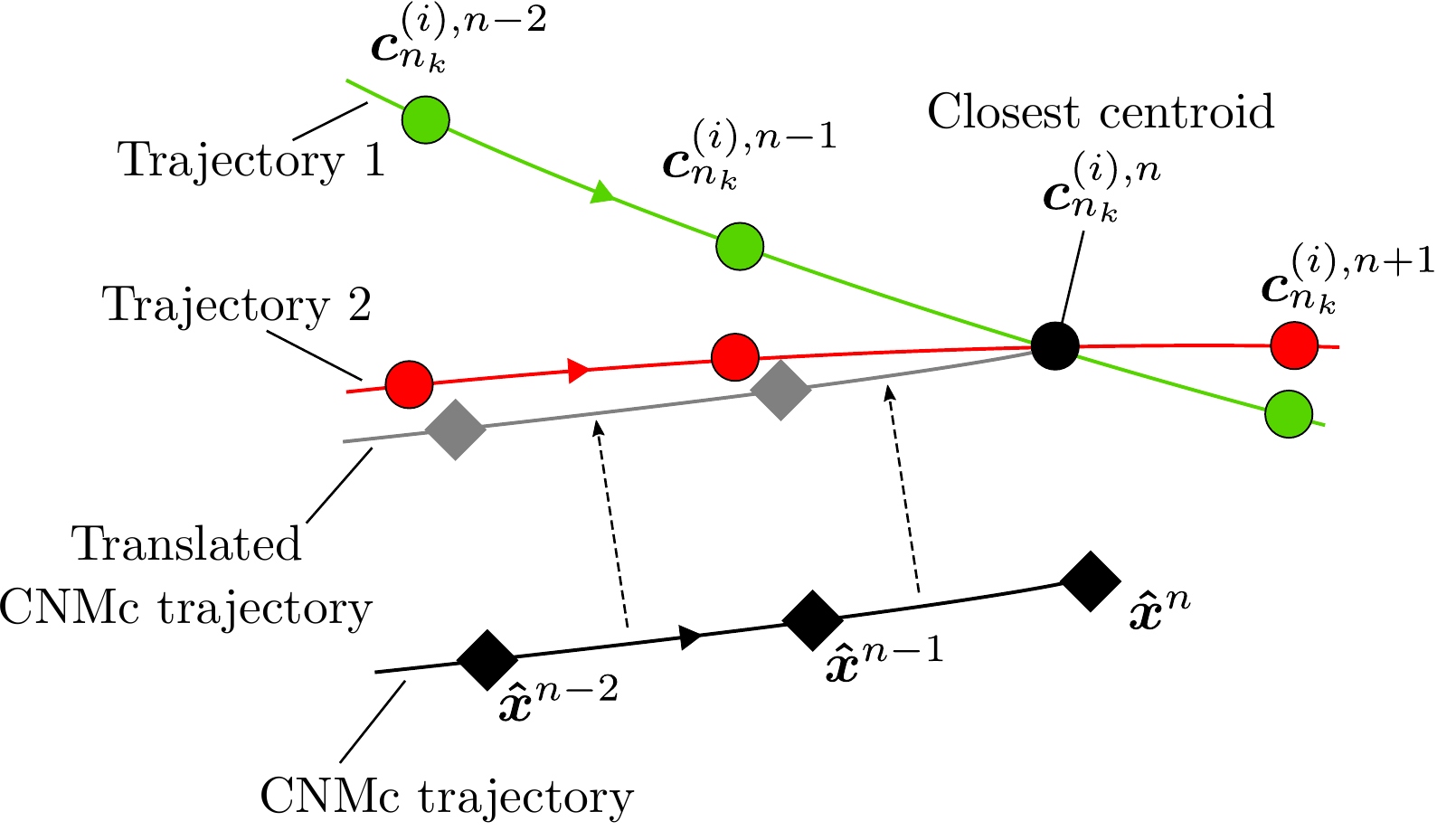}
      \end{center}
      \caption{
        \textbf{Determination of the appropriate past trajectory.}
        Example of a neighboring centroid $\bm{c}_{n_k}^{(i),n}$ to the predicted state
        $\hat{\bm{x}}$ with two possible past trajectories.
The goal is to determine which of trajectories 1 or 2 is the most
        similar to the CNMc trajectory.
        First, the CNMc trajectory is translated so that $\hat{\bm{x}}^n$
        coincides with $\bm{c}_{n_k}^{(i),n}$. Then, the past predicted states
        $\hat{\bm{x}}^{n-l}$ are sequentially compared to the previous centroids
        $\bm{c}_{n_k}^{(i),n-l}$ of trajectories 1 and 2, $l=1,\ldots,L-1$.
        The trajectory with the smallest difference is selected as past for
        $\bm{c}_{n_k}^{(i),n}$.
        In this example, trajectory 2 is selected, since it is most
        aligned with the CNMc trajectory.
      }
      \label{Fig:PastTraj}
    \end{figure}

  \item[Step 3:] The transition from $\hat{\bm{x}}^{n}$ to $\hat{\bm{x}}^{n+1}$
    is fully characterized by the transition vector $\hat{\bm{v}}$ and the
    transition time $\hat{T}$, computed by averaging those of the $2N_k$ nearest
    centroids $\bm{c}_{n_k}^{(i),n}$. A weighted average can be employed to
    account for the distance from $\hat{\bm{b}}$ to $\bm{b}^{(1)}$ and
    $\bm{b}^{(2)}$. In the case where $\hat{\bm{b}}$ is equally distant to
    $\bm{b}^{(1)}$ and $\bm{b}^{(2)}$, $\hat{T}$ and $\hat{\bm{v}}$ are computed
    as
    \begin{align}
      \hat{T} & = \frac{1}{2N_k}\sum_{i=1}^{2}\sum_{n_k=1}^{N_{k}}T_{n_k}^{(i)}\label{Eq:TAv}\,,\\
      \hat{\bm{v}} & = \frac{1}{2N_k}\sum_{i=1}^{2}\sum_{n_k=1}^{N_k}\bm{v}_{n_k}^{(i)}\,.\label{Eq:vAv}
    \end{align}
The new position is $\hat{\bm{x}}^{n+1}=\hat{\bm{x}}^{n}+\hat{\bm{v}}$ and
    is reached in a corresponding time of $t^{n+1}=t^{n}+\hat{T}$.
    
  \item[Step 4:] The optional final step is the position correction. Observations  have
    shown that the predicted dynamics sometimes tend to slide toward one of the
    two neighboring operating conditions.
To circumvent this issue, the position correction forces the state to have a
    constant relative distance between the two test cases (1) and (2).
The distance $d^{(i)}$ between the predicted state $\hat{\bm{x}}$ and a
    neighboring test case
    $(i)$ is defined as the distance between $\hat{\bm{x}}$ and the closest
    snapshot in this test case $\bm{x}^{(i)}$ as
\begin{equation}
d^{(i)}=d\left(\hat{\bm{x}},\bm{x}^{(i)}\right) =
      \lVert\hat{\bm{x}}-\bm{x}^{(i)}\rVert\,.
    \end{equation}
The correction is formulated to ensure that the ratio $r$ of the distances between the corrected position
    $\hat{\bm{x}}_{\text{corr}}^{n+1}$ and the two neighboring test cases is the same as
    that of the distances between $\hat{\bm{b}}$ and the two neighboring control
    inputs
\begin{equation}
      \label{}
      r = \frac{d\left(\hat{\bm{b}},\bm{b}^{(1)}\right)}{d\left(\hat{\bm{b}},\bm{b}^{(2)}\right)}
      = \frac{
          d\left(\hat{\bm{x}}_{\text{corr}}^{n+1},\bm{x}^{(1)}\right)
        }{
          d\left(\hat{\bm{x}}_{\text{corr}}^{n+1},\bm{x}^{(2)}\right)
        }\,.
    \end{equation}
This is achieved by computing $\hat{\bm{x}}_{\text{corr}}^{n+1}$ as
\begin{equation}
      \label{Eq:PosCorr}
      \hat{\bm{x}}_{\text{corr}}^{n+1}=
      \frac{
        d_{\bm{b}}^{(2)}\bm{x}^{(1)} +
        d_{\bm{b}}^{(1)}\bm{x}^{(2)}
      }
      {
        d_{\bm{b}}^{(1)} + d_{\bm{b}}^{(2)}
      }\,,
    \end{equation}
where $d_{\bm{b}}^{(i)}=d(\hat{\bm{b}},\bm{b}^{(i)})$ is given by
    (\ref{Eq:Distb}).
\end{description}

\begin{figure}[h!]
  \begin{center}
    \includegraphics[width=11.5cm]{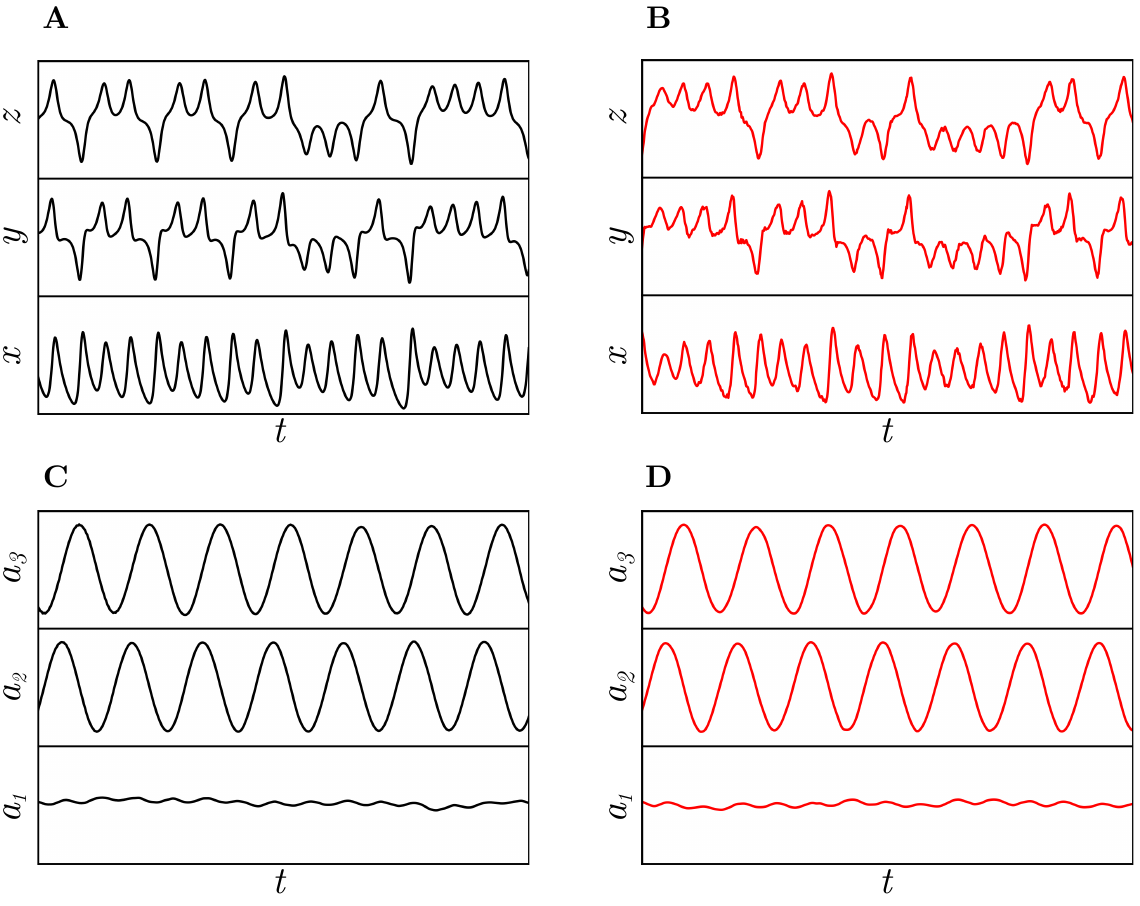}
  \end{center}
  \caption{
    \textbf{Time series modeling with CNMc.}
    Time series of the reference data (black) and the predicted cases (red) for
    the (\textbf{A}) (\textbf{B}) the Lorenz system and the (\textbf{C}) (\textbf{D}) actuated
    turbulent boundary layer.
    The Lorenz attractor with $\rho=28$ is interpolated from two test cases with
    $\rho = 26$ and $\rho=30$ and the boundary layer with actuation parameters
    $\lambda^+=1000$, $T^+=120$ and $A^+=30$ is interpolated from cases with
    $\lambda^+=1000$, $T^+=120$, $A^+=20$, and
    $\lambda^+=1000$, $T^+=120$ and $A^+=40$.
For both applications, the main dynamical features are well reconstructed.
  }
  \label{Fig:CNMc:TS}
\end{figure}

In addition to the two CNM parameters (number of centroids $K$ and model order
$L$), CNMc requires three more settings: The number of closest centroids in
each neighboring test case, the norm for the nearest neighbor distance and the
optional position correction.
These parameters can be optimized for each application separately.
One possible approach to optimize these parameters is the hold-out method, where
the RMSE is evaluated on the autocorrelation functions.
The CNMc parameters for the interpolated Lorenz system and for the 
boundary layer are summarized in Table \ref{Tab:CNMcSettings}.
The corresponding time series for these two applications are presented in
Fig.~\ref{Fig:CNMc:TS}.
\begin{table}[h!]
  \centering
  \caption{\textbf{CNMc settings for the Lorenz system and boundary layer
  applications.}}
  \begin{tabular}{>{\centering}m{2.6cm}|>{\centering}m{2cm}>{\centering}m{1.8cm}>{\centering}m{2.5cm}>{\centering}m{2.5cm}>{\centering\arraybackslash}m{2cm}}
\textbf{Application} & \textbf{Number of clusters $\bm{K}$} & \textbf{Model
      order $\bm{L}$} & \textbf{Number of neighboring centroids} &
      \textbf{Distance norm} & \textbf{Position correction} \\
    \toprule
    Lorenz system  & 250 & 83 & 1 & Euclidean & No \\
    Boundary layer & 120 & 15 & 3 & $L_{10}$  & Yes\\
  \end{tabular}
  \label{Tab:CNMcSettings}
\end{table}

\begin{algorithm}[H]
\SetAlgoLined
Extract the centroids of the two closest test cases\;
Initialize the state $\hat{\bm{x}}^{n=0}=\hat{\bm{x}}^0$\;
Time initialization $t^{n=0}=t^{0}=0$\;
\While{$t^{n}<T_{max}$}{
    \For{$i \leftarrow 1,2$}{
      Find the $N_k$ nearest centroids from test case $(i)$: $\bm{c}_{n_k}^{(i),n}$, $n_k=1,\ldots,N_k$\;
      \For{$n_k\leftarrow 1$ \KwTo $N_k$}{
        Identify past trajectory of $\bm{c}_{n_k}^{(i),n}$\;
        Get transition time $T_{n_k}^{(i)}$\; Get next centroid $\bm{c}_{n_k,\text{next}}^{(i)}$, and transition vector $\bm{v}_{n_k}^{(i)}=\bm{c}_{n_k}^{(i),n+1}-\bm{c}_{n_k}^{(i),n}$\;
        }
      }
      Average transition time $\hat{T} = \frac{1}{2N_k}\sum_{i=1}^{2}\sum_{n_k=1}^{N_{k}}T_{n_k}^{(i)}$\;

      Average transition vector $\hat{\bm{v}} = \frac{1}{2N_k}\sum_{i=1}^{2}\sum_{n_k=1}^{N_k}\bm{v}_{n_k}^{(i)}$\;

      Update position $\hat{\bm{x}}^{n+1}=\hat{\bm{x}}^{n}+\hat{\bm{v}}$\;
      Correct position $\hat{\bm{x}}_{\text{corr}}^{n+1}$\;
      Update time $t^{n+1}=t^n+\hat{T}$\;
      Update iteration number $n=n+1$
    }

\caption{
  CNMc procedure
  \label{Algo:CNMc}
}
\end{algorithm}

 \end{appendices}

\bibliographystyle{ScienceAdvances}
\bibliography{Main}

\begin{thebibliography}{10}

\bibitem{Holmes2012Book}
P.~Holmes, J.~L. Lumley, G.~Berkooz, C.~W. Rowley, {\it Turbulence, coherent
  structures, dynamical systems and symmetry\/} (Cambridge university press,
  2012).

\bibitem{Benner2015SIAM}
P.~Benner, S.~Gugercin, K.~Willcox, A survey of projection-based model
  reduction methods for parametric dynamical systems.
\newblock {\it SIAM Review\/} {\bf 57}, 483--531 (2015).

\bibitem{Tu2014JCD}
J.~H. Tu, C.~W. Rowley, D.~M. Luchtenburg, S.~L. Brunton, J.~N. Kutz, On
  dynamic mode decomposition: {Theory} and applications.
\newblock {\it Journal of Computational Dynamics\/} {\bf 1}, 391 (2014).

\bibitem{Ye2015PNAS}
H.~Ye, R.~J. Beamish, S.~M. Glaser, S.~C. Grant, C.-h. Hsieh, L.~J. Richards,
  J.~T. Schnute, G.~Sugihara, Equation-free mechanistic ecosystem forecasting
  using empirical dynamic modeling.
\newblock {\it Proceedings of the National Academy of Sciences\/} {\bf 112},
  E1569--E1576 (2015).

\bibitem{Brunton2020arfm}
S.~L. Brunton, B.~R. Noack, P.~Koumoutsakos, Machine learning for fluid
  mechanics.
\newblock {\it Ann.~Rev.~Fluid Mech.\/} {\bf 52}, 477--508 (2020).

\bibitem{Bongard2007PNAS}
J.~Bongard, H.~Lipson, Automated reverse engineering of nonlinear dynamical
  systems.
\newblock {\it Proceedings of the National Academy of Sciences\/} {\bf 104},
  9943--9948 (2007).

\bibitem{Schmidt2009science}
M.~Schmidt, H.~Lipson, Distilling {Free}-{Form} {Natural} {Laws} from
  {Experimental} {Data}.
\newblock {\it Science\/} {\bf 324}, 81--85 (2009).

\bibitem{Brunton2016PNAS}
S.~L. Brunton, J.~L. Proctor, J.~N. Kutz, Discovering governing equations from
  data by sparse identification of nonlinear dynamical systems.
\newblock {\it Proceedings of the national academy of sciences\/} {\bf 113},
  3932--3937 (2016).

\bibitem{Fu1973IJC}
F.~C. Fu, J.~B. Farison, On the {Volterra} series functional evaluation of the
  response of non-linear discrete-time systems.
\newblock {\it International Journal of Control\/} {\bf 18}, 553--558 (1973).

\bibitem{Chatfield2000Book}
C.~Chatfield, {\it Time-series forecasting\/} (CRC press, 2000).

\bibitem{Juang1994Book}
J.-N. Juang, {\it Applied system identification\/} (Prentice-Hall, Inc., 1994).

\bibitem{Wang2015IEEE}
T.~Wang, H.~Gao, J.~Qiu, A combined adaptive neural network and nonlinear model
  predictive control for multirate networked industrial process control.
\newblock {\it IEEE Transactions on Neural Networks and Learning Systems\/}
  {\bf 27}, 416--425 (2016).

\bibitem{Newman2008PT}
M.~Newman, The physics of networks.
\newblock {\it Physics today\/} {\bf 61}, 33--38 (2008).

\bibitem{Barabasi1999Science}
A.-L. Barabási, R.~Albert, Emergence of scaling in random networks.
\newblock {\it Science\/} {\bf 286}, 509--512 (1999).

\bibitem{Barabasi2003SA}
A.-L. Barabási, E.~Bonabeau, Scale-free networks.
\newblock {\it Scientific American\/} {\bf 288}, 60--69 (2003).

\bibitem{Norris1998}
J.~R. Norris, {\it Markov chains\/}, no.~2 (Cambridge university press, 1998).

\bibitem{Marwan2009PLA}
N.~Marwan, J.~F. Donges, Y.~Zou, R.~V. Donner, J.~Kurths, Complex network
  approach for recurrence analysis of time series.
\newblock {\it Physics Letters A\/} {\bf 373}, 4246--4254 (2009).

\bibitem{Taira2016JFM}
K.~Taira, A.~G. Nair, S.~L. Brunton, Network structure of two-dimensional
  decaying isotropic turbulence.
\newblock {\it Journal of Fluid Mechanics\/} {\bf 795} (2016).

\bibitem{Kaiser2014JFM}
E.~Kaiser, B.~R. Noack, L.~Cordier, A.~Spohn, M.~Segond, M.~Abel, G.~Daviller,
  J.~Östh, S.~Krajnović, R.~K. Niven, Cluster-based reduced-order modelling
  of a mixing layer.
\newblock {\it Journal of Fluid Mechanics\/} {\bf 754}, 365--414 (2014).

\bibitem{Li2020JFM}
H.~Li, D.~Fernex, R.~Semaan, J.~Tan, M.~Morzyński, B.~R. Noack, Cluster-based
  network model.
\newblock {\it Journal of Fluid Mechanics\/} {\bf {\rm (in print), see
  arXiv:2001.02911}} (2020).

\bibitem{Ching2013Book}
W.-K. Ching, X.~Huang, M.~K. Ng, T.-K. Siu, {\it Markov {Chains}: {Models},
  {Algorithms} and {Applications}\/}, International {Series} in {Operations}
  {Research} \& {Management} {Science} (Springer, 2013), pp. 141--176.

\bibitem{Daniels2015NC}
B.~C. Daniels, I.~Nemenman, Automated adaptive inference of phenomenological
  dynamical models.
\newblock {\it Nature Communications\/} {\bf 6}, 8133 (2015).

\bibitem{David2006}
D.~Arthur, S.~Vassilvitskii, k-means++: {The} advantages of careful seeding,
  {\it Tech. rep.\/}, Stanford (2006).

\bibitem{Jain1999ACMCS}
A.~K. Jain, M.~N. Murty, P.~J. Flynn, Data clustering: a review.
\newblock {\it ACM Computing Surveys\/} {\bf 31}, 264--323 (1999).

\bibitem{Takens1981}
F.~Takens, {\it Dynamical systems and turbulence, {Warwick} 1980\/}, Lecture
  {Notes} in {Mathematics} (Springer, University of Warwick, 1981), pp.
  366--381.

\bibitem{Lorenz1963JAS}
E.~N. Lorenz, Deterministic {Nonperiodic} {Flow}.
\newblock {\it Journal of the Atmospheric Sciences\/} {\bf 20}, 130--141
  (1963).

\bibitem{Protas2015JFM}
B.~Protas, B.~R. Noack, J.~Östh, Optimal nonlinear eddy viscosity in
  {Galerkin} models of turbulent flows.
\newblock {\it Journal of Fluid Mechanics\/} {\bf 766}, 337--367 (2015).

\bibitem{Rossler1976PLA}
O.~E. Rössler, An equation for continuous chaos.
\newblock {\it Physics Letters A\/} {\bf 57}, 397--398 (1976).

\bibitem{Brunton2017NC}
S.~L. Brunton, B.~W. Brunton, J.~L. Proctor, E.~Kaiser, J.~N. Kutz, Chaos as an
  intermittently forced linear system.
\newblock {\it Nature Communications\/} {\bf 8}, 19 (2017).

\bibitem{Farazmand2017SA}
M.~Farazmand, T.~P. Sapsis, A variational approach to probing extreme events in
  turbulent dynamical systems.
\newblock {\it Science Advances\/} {\bf 3}, e1701533 (2017).

\bibitem{Albers2020FTaC}
M.~Albers, P.~S. Meysonnat, D.~Fernex, R.~Semaan, B.~R. Noack, W.~Schröder,
  Drag {Reduction} and {Energy} {Saving} by {Spanwise} {Traveling}
  {Transversal} {Surface} {Waves} for {Flat} {Plate} {Flow}.
\newblock {\it Flow, Turbulence and Combustion\/} {\bf 105}, 125--157 (2020).

\bibitem{Wan2018PLOS}
Z.~Y. Wan, P.~Vlachas, P.~Koumoutsakos, T.~Sapsis, Data-assisted reduced-order
  modeling of extreme events in complex dynamical systems.
\newblock {\it PLOS ONE\/} {\bf 13}, e0197704 (2018).

\bibitem{Rossum2011Book}
G.~Van~Rossum, F.~L. Drake, {\it The python language reference manual\/}
  (Network Theory Ltd., 2011).

\bibitem{Pauli2020Nature}
P.~Virtanen, R.~Gommers, T.~E. Oliphant, M.~Haberland, T.~Reddy, D.~Cournapeau,
  E.~Burovski, P.~Peterson, W.~Weckesser, J.~Bright, S.~J. van~der Walt,
  M.~Brett, J.~Wilson, K.~J. Millman, N.~Mayorov, A.~R.~J. Nelson, E.~Jones,
  R.~Kern, E.~Larson, C.~J. Carey, I.~Polat, Y.~Feng, E.~W. Moore,
  J.~VanderPlas, D.~Laxalde, J.~Perktold, R.~Cimrman, I.~Henriksen, E.~A.
  Quintero, C.~R. Harris, A.~M. Archibald, A.~H. Ribeiro, F.~Pedregosa, P.~van
  Mulbregt, {SciPy} 1.0: fundamental algorithms for scientific computing in
  {Python}.
\newblock {\it Nature Methods\/} {\bf 17}, 261--272 (2020).

\bibitem{Goldberger2000C}
A.~L. Goldberger, L.~A. Amaral, L.~Glass, J.~M. Hausdorff, P.~C. Ivanov, R.~G.
  Mark, J.~E. Mietus, G.~B. Moody, C.~K. Peng, H.~E. Stanley, {PhysioBank},
  {PhysioToolkit}, and {PhysioNet}: components of a new research resource for
  complex physiologic signals.
\newblock {\it Circulation\/} {\bf 101}, 215--220 (2000).

\bibitem{Fylliaditakis2018JAMP}
E.~D. Fylladitakis, Kolmogorov {Flow}: {Seven} {Decades} of {History}.
\newblock {\it Journal of Applied Mathematics and Physics\/} {\bf 6},
  2227--2263 (2018).

\bibitem{Farazmand2016PRE}
M.~Farazmand, T.~P. Sapsis, Dynamical indicators for the prediction of bursting
  phenomena in high-dimensional systems.
\newblock {\it Physical Review E\/} {\bf 94}, 032212 (2016).

\bibitem{Bechert1985AIAA}
D.~Bechert, W.~Reif, {\it 23rd {Aerospace} {Sciences} {Meeting}\/} (American
  Institute of Aeronautics and Astronautics, 1985), p. 546.

\bibitem{Luhar2016JoT}
M.~Luhar, A.~S. Sharma, B.~J. McKeon, On the design of optimal compliant walls
  for turbulence control.
\newblock {\it Journal of Turbulence\/} {\bf 17}, 787--806 (2016).

\bibitem{Du200Science}
Y.~Du, G.~E. Karniadakis, Suppressing wall turbulence by means of a transverse
  traveling wave.
\newblock {\it Science\/} {\bf 288}, 1230--1234 (2000).

\bibitem{Quadrio2011PTRSA}
M.~Quadrio, Drag reduction in turbulent boundary layers by in-plane wall
  motion.
\newblock {\it Philosophical Transactions of the Royal Society A: Mathematical,
  Physical and Engineering Sciences\/} {\bf 369}, 1428--1442 (2011).

\bibitem{Fernex2020PRF}
D.~Fernex, R.~Semaan, M.~Albers, P.~S. Meysonnat, W.~Schröder, B.~R. Noack,
  Actuation response model from sparse data for wall turbulence drag reduction.
\newblock {\it Physical Review Fluids\/} {\bf 5}, 073901 (2020).

\bibitem{MacQueen1967Proceeding}
J.~MacQueen (University of California, 1967), vol.~1. Conference Name:
  Proceedings of the Fifth Berkeley Symposium on Mathematical Statistics and
  Probability.

\bibitem{Lloyd1982IEEE}
S.~Lloyd, Least squares quantization in {PCM}.
\newblock {\it IEEE Transactions on Information Theory\/} {\bf 28}, 129--137
  (1982). Conference Name: IEEE Transactions on Information Theory.

\bibitem{Fernex2019PAMM}
D.~Fernex, R.~Semaan, M.~Albers, P.~S. Meysonnat, W.~Schröder, R.~Ishar,
  E.~Kaiser, B.~R. Noack, Cluster-based network model for drag reduction
  mechanisms of an actuated turbulent boundary layer.
\newblock {\it Proceedings in Applied Mathematics and Mechanics\/} {\bf 19}
  (2019).

\bibitem{Cao2014EF}
Y.~Cao, E.~Kaiser, J.~Borée, B.~R. Noack, L.~Thomas, S.~Guilain, Cluster-based
  analysis of cycle-to-cycle variations: application to internal combustion
  engines.
\newblock {\it Experiments in Fluids\/} {\bf 55}, 1837 (2014).

\bibitem{Ishar2019JFM}
R.~Ishar, E.~Kaiser, M.~Morzyński, D.~Fernex, R.~Semaan, M.~Albers, P.~S.
  Meysonnat, W.~Schröder, B.~R. Noack, Metric for attractor overlap.
\newblock {\it Journal of Fluid Mechanics\/} {\bf 874}, 720--755 (2019).

\bibitem{Osth2015JWEIA}
J.~Östh, E.~Kaiser, S.~Krajnović, B.~R. Noack, Cluster-based reduced-order
  modelling of the flow in the wake of a high speed train.
\newblock {\it Journal of Wind Engineering and Industrial Aerodynamics\/} {\bf
  145}, 327--338 (2015).

\bibitem{Bentley1975ACM}
J.~L. Bentley, Multidimensional binary search trees used for associative
  searching.
\newblock {\it Communications of the ACM\/} {\bf 18}, 509--517 (1975).

\end{thebibliography}

\end{document}